\documentclass[
11pt,
showpacs, showkeys,
nofootinbib,
 amsmath,amssymb,
 aps,
pre,
floatfix,
]{revtex4-1}

\pdfoutput=1
\usepackage{graphicx}

\usepackage{dcolumn}
\usepackage{bm}
\usepackage{hyperref}

\usepackage{subfigure}
\usepackage{longtable}
\usepackage{lipsum}
\usepackage{color}
\usepackage{comment}

\usepackage{amsmath}
\usepackage{amssymb}

\newcommand{\ElectricPotential}{{\Phi}}
\newcommand{\VolumeFraction}{{\phi}}
\newcommand{\SpecificCharge}{{z}}

\newcommand{\SpeciesFlux}{\boldsymbol{\mathcal{F}}}

\newcommand{\DiffusionMatrix}{\chi}
\newcommand{\MaxwellTensor}{\boldsymbol{\sigma}}
\newcommand{\ViscousTensor}{\boldsymbol{\tau}}

\newcommand{\half}{\frac{1}{2}}

\global\long\def\V#1{\boldsymbol{#1}}
\global\long\def\M#1{\boldsymbol{#1}}

\global\long\def\grad{\M{\nabla}}

\newcommand{\modified}[1]{#1}
\newcommand{\deleted}[1]{}
\newcommand{\fixed}[1]{#1}
\newcommand{\new}[1]{#1}

\begin{document}

\title{Low Mach Number Fluctuating Hydrodynamics for Electrolytes}

\author{Jean-Philippe P\'eraud$^{1}$, Andy Nonaka$^{1}$, Anuj Chaudhri$^{1}$,
John B. Bell$^{1}$, Aleksandar Donev$^{2}$, and Alejandro L. Garcia$^{3}$ }
\affiliation{$^1$ Computational Research Division, Lawrence Berkeley National Laboratory \\
 1 Cyclotron Road, Berkeley, CA 94720 \\ }
\affiliation{$^2$ Courant Institute of Mathematical Sciences, New York University \\
 251 Mercer Street, New York, NY 10012 \\ }
\affiliation{$^3$ Department of Physics and Astronomy, San Jose State University \\
 1 Washington Square, San Jose, CA 95192 \\ }

\date{\today}

\begin{abstract}

We formulate and study computationally the low Mach number fluctuating hydrodynamic
equations for electrolyte solutions.
We are interested in studying transport in mixtures of charged species
at the mesoscale, down to scales below the Debye length,
where thermal fluctuations have a significant impact on the dynamics.
Continuing our previous work on fluctuating hydrodynamics of multicomponent mixtures of
incompressible isothermal miscible liquids (A. Donev, et al., Physics of Fluids, 27, 3, 2015),
we now include the effect of charged species using a quasielectrostatic approximation.
Localized charges create an electric field, which in turn provides
additional forcing in the mass and momentum equations.
Our low Mach number formulation
eliminates sound waves from the fully compressible formulation and leads to a
more computationally efficient quasi-incompressible formulation.  We demonstrate
our ability to model saltwater (NaCl) solutions in both equilibrium
and nonequilibrium settings.  We show that our algorithm is second-order in the deterministic
setting, and for length scales much greater than the Debye length
gives results consistent with an electroneutral/ambipolar approximation.
In the stochastic setting, our model captures the predicted dynamics of
equilibrium and nonequilibrium
fluctuations.  We also identify and model an instability
that appears when diffusive mixing occurs in the presence of an applied electric field.

\end{abstract}

\pacs{05.40.-a, 47.11.-j, 47.10.ad, 47.11.St, 47.55.pd, 47.65.-d}

\keywords{fluctuating hydrodynamics,
          computational fluid dynamics,
          Navier-Stokes equations,
          low Mach number methods,
          multicomponent diffusion,
          electrohydrodynamics}

\maketitle

\section{Introduction}

At macroscopic scales, fluid dynamics is governed by partial differential equations
that characterize the behavior of the fluid in terms of smoothly evolving fields
that represent density, momentum, and other characteristics of the fluid.  However,
at atomic scale fluids are discrete systems composed of individual molecules whose
dynamics are governed by complex interaction potentials.
The discrepancy between these two descriptions is manifest at the mesoscale.
While it is possible to model a fluid using macroscopic field variables at the mesoscale,
we know that they are no longer smooth fields; instead they fluctuate even for systems
that are at thermodynamic equilibrium.

Fluctuations in systems at equilibrium are well understood;
for systems that are not close to a critical point, basic statistical mechanics
provides a complete characterization.
In this setting fluctuations are benign; they are simply small stochastic
variations about the mean behavior.  However, in systems that are out of equilibrium,
fluctuations can have a significant impact on macroscopic behavior.
A macroscopic gradient is sufficient to significantly affect
the mesoscale dynamics as manifested by the enhancement of fluctuations,
both in magnitude and their range of influence. In addition, some quantities
that are uncorrelated at equilibrium (e.g., fluctuations of concentration
and fluid velocity) are found to be correlated in non-equilibrium systems;
these correlations can produce macroscopic effects, such as the
``giant fluctuation'' phenomena, which are observed
in laboratory experiments~\cite{GiantFluctuations_Nature,GiantFluctuations_Summary}.

In principle the effects of fluctuations can be studied using
all-atom molecular simulations. In practice, however, this type of microscopic modeling
will be infeasible, even on proposed exascale architectures,
for many mesoscopic problems of interest.
A more efficient and tractable numerical approach for mesoscopic fluids is
fluctuating hydrodynamics~\cite{Landau:Fluid,FluctHydroNonEq_Book}.
This theory extends conventional hydrodynamics by
including a random component to the dissipative fluxes.
The form of these stochastic fluxes is obtained from irreversible thermodynamics
and the fluctuation-dissipation theorem.

In a series of papers, we developed numerical methodology for fluctuating hydrodynamics of
multicomponent mixtures of compressible fluids \cite{MultispeciesCompressible} and
quasi-incompressible miscible liquids
\cite{LowMachExplicit,LowMachImplicit,LowMachMultispecies}.
We used low Mach number asymptotics to derive an alternative set of hydrodynamic
equations that do not contain fast acoustic waves.  For flows in the low Mach number regime,
where the characteristic fluid velocity is small compared to the sound speed
($U \lesssim 0.1 c$), sound
waves are sufficiently weak that they do not change the thermodynamics of the system.
Thus, for these classes of problems a low Mach number approach can be an order
of magnitude or more computationally efficient than a fully compressible approach.
Our method correctly captures the predicted dynamics of equilibrium and nonequilibrium
fluctuations, and is able to
model experimentally observed phenomena such as mixed-mode instability and
diffusive layer convection between layers of salt solution and sucrose \cite{MixedDiffusiveInstability,LowMachMultispecies}.

In this paper, we extend our multicomponent fluctuating hydrodynamics 
approach \cite{LowMachMultispecies} to include charged species. 
Transport phenomena in electrolytes are important both for studying naturally occurring and man-made systems.
In living cells this is of particular interest since transport is known to rely strongly
on membrane potentials and the electrodiffusion of ions \cite{grodzinsky2011field}.
Being able to model such systems with the
inclusion of their inherent statistical fluctuations would not only be a way to increase our understanding
of cellular mechanisms, but also provides a path towards better modeling tools for bio-engineering
applications. Fields such as microfabrication would also benefit from such numerical tools. For instance,
the synthesis of nanowires often relies on electrodeposition processes and, while the deposition techniques
by themselves are well developed, non-homogeneous growth rates induced by random fluctuations have been reported
\cite{bograchev2013nanowires}. Batteries and fuel cells are another example of applications relying on ionic transport.
In all these examples the length and time scales involved are usually intractable for direct
modeling methods such as molecular dynamics. By contrast, fluctuating hydrodynamics provides a naturally-suited framework. And while there are alternative numerical approaches
(e.g., lattice Boltzmann for fluctuating hydrodynamics~\cite{LB_StatMech}
and for electrolytes~\cite{LatticeBoltzElectrolytes}) the methodology presented here
is based on well-established schemes in computational fluid dynamics.

In this paper we
model strong electrolyte solutions, such as salt (NaCl) dissolved in water.
The electric field, \modified{resulting from an applied field} and internal free charges, acts
upon the charged species resulting in additional forcing in the mass and momentum equations.
We consider isothermal systems and neglect magnetic
effects by using the quasielectrostatic approximation.
\fixed{Dreyer {\em et al.} \cite{NernstPlanckModel} have presented a closely-related deterministic
formulation (which, like our formulation, is fully consistent with
nonequilibrium thermodynamics \cite{DM_63}) of the complete hydrodynamic equations for an ideal ternary mixture 
containing a neutral solvent.}
Our formulation does not assume ideality and treats all species on an equal footing, allowing for the modeling
of mixed solvents (e.g., ethanol and water) with arbitrary mixtures of solute ions of differing valences.
In modeling the transport of ions in electrolytes it is often assumed that
the solution is locally neutral, as in the Nernst-Hartley theory of diffusion~\cite{robinson2012electrolyte,grodzinsky2011field}.
Since we are interested here in resolving scales comparable and even smaller than the Debye length,
local electroneutrality is \emph{not} imposed in our model since,
at such small scales, there are significant fluctuations in the total charge density.
In this paper we focus on hydrodynamic transport so chemical reactions
(e.g., disassociation and recombination in weak electrolytes) are omitted;
see~\cite{MultispeciesChemistry} for a discussion of how to include chemistry.

This paper is divided into the following sections:
In Section \ref{ModelSection}, we show how
the presence of charged species and, therefore, of a non-zero electric field
modifies the multispecies fluctuating transport equations.
In Section \ref{sec:Structure}, we discuss the structure factor calculations \cite{LLNS_S_k}
that we later use to validate the algorithm.
In Section \ref{NumericsSection}, we describe a numerical scheme for solving the resulting
equations.
In Section \ref{ExamplesSection}, we provide numerical examples intended to both verify the
correctness and accuracy of the code.
In particular, we apply the code to a model of seawater and
first check that the code is second-order in space and time in the deterministic setting.
Next, we show that our algorithm correctly captures equilibrium fluctuations
by calculating the associated structure factors.
We then study the phenomenon of giant fluctuations that emerge in the presence of an
imposed concentration gradient.  We finally observe the mixing instability that emerges in a
interfacial mixing system subjected to a potential gradient normal to the interface.

\section{Fluctuating Hydrodynamics for Electrolytes}\label{ModelSection}

In this section we present our model equations for multicomponent electrolyte fluids.
\modified{We consider a system consisting of a fluid mixture of neutral and charged species}.
\deleted{separating the contributions of the long-ranged Coulomb electrostatic interactions from those due to short-ranged molecular forces.}
We define component mass densities, $\V{\rho} = (\rho_1,\ldots,\rho_N)$, with total mass
density $\rho = \sum_{k=1}^N \rho_k$ and mass fractions
$\V{w} = (w_1,\ldots,w_N) = {\rho}^{-1}(\rho_1,\ldots,\rho_N)$.
The mole fractions can be expressed in terms of mass fraction as
$\V{x} = (x_1,\ldots,x_N) = \bar{m}(\frac{w_1}{m_1},\ldots,\frac{w_N}{m_N})$,
where $m_k$ is the mass of a molecule of species $k$ and
\begin{equation}
\bar{m} = \left( \sum_{k=1}^{N} \frac{w_k}{m_k} \right)^{-1}
\end{equation}
is the mixture-averaged molecular mass.
Note that $n = \rho/\bar{m}$ is the total number density.

In the presence of charges, we denote the charge per unit mass for each component as
$\V{\SpecificCharge} = (\SpecificCharge_1,\ldots,\SpecificCharge_N)$.
Thus, the component charge density for free charges
is $q_k^\mathrm{f} = \rho_k \SpecificCharge_k$ and
$q^\mathrm{f} = \sum_{k=1}^N q_k^\mathrm{f}$ is the total charge density for free charges.
Note that $\SpecificCharge_k = Q_k F / (m_k N_a)$, where
$Q_k$ is the valence of species $k$, $F$ is Faraday's constant,
and $N_a$ is Avogadro's number.

\subsection{Mass Transport}

From our previous work on neutral multicomponent transport
\cite{LowMachMultispecies}, the evolution of the mass densities $\rho_k(\V r,t)$ is given by
\begin{equation}
\frac{\partial\rho_k}{\partial t} + \nabla\cdot (\rho_k \V{v}) = -\nabla\cdot \SpeciesFlux_k,\label{SpeciesTransportEqn}
\end{equation}
where $\V{v}(\V r,t)$ is the fluid velocity and the $\SpeciesFlux_k$ are the barycentric species fluxes.
Note that by summing (\ref{SpeciesTransportEqn}) over species we obtain the continuity equation,
\begin{equation}\label{ContinuityEqn}
\frac{\partial\rho}{\partial t} + \nabla\cdot (\rho \V{v}) = 0,
\end{equation}
since $\sum_{k=1}^N \SpeciesFlux_k = 0$.  In fluctuating hydrodynamics the barycentric species
flux has two contributions,
$\SpeciesFlux_k = \overline{\SpeciesFlux}_k + \widetilde{\SpeciesFlux}_k$,
which are the deterministic flux, $\overline{\SpeciesFlux}_k$, and
the stochastic flux, $\widetilde{\SpeciesFlux}_k$.
The stochastic flux is a mean zero Gaussian random field that generates
fluctuations of the mass densities.

\subsubsection{General Formulation}

From \cite{LowMachMultispecies},
the deterministic fluxes are given by
\begin{equation}\label{FickianFluxEqn}
\overline{\SpeciesFlux}_k = -\rho w_k\sum_{j=1}^{N}
\DiffusionMatrix_{k,j} \left( \V{d}_{j} + \frac{\zeta_{j}}{T}\nabla T \right),
\end{equation}
where $\V{\DiffusionMatrix}$ is the symmetric multicomponent diffusion matrix,
the $\V{d}_{j}$ are the thermodynamic driving forces,
$\zeta_j$ are the thermal diffusion ratios and
$T$ is the temperature.
This flux is formulated here in its Fickian
form however, as outlined in \cite{LowMachMultispecies},
the diffusion matrix $\V{\DiffusionMatrix}$ is best obtained by way of
Maxwell-Stefan (MS) theory~\cite{MaxStefElectrolytes}.
Note that mass diffusion in electrolytes is qualitatively different
than in neutral mixtures (e.g.,
see~\cite{NegativeMaxStefanDiff1,NegativeMaxStefanDiff2,DivergentMaxStefanDiff})
and in Section~\ref{sec:Electroneutral} we discuss some of the commonly used approximations.
The MS diffusion coefficients can in principle be obtained from molecular dynamics
simulations, however,
unlike for uncharged mixtures, they show rather strong dependence on concentration \cite{MS_diffusion_ionic}.

For neutral fluids the thermodynamics driving forces
include contributions from compositional
gradients and barodiffusion (pressure gradients). When charges are included, the driving force also includes
an electrostatic term so that
\begin{equation}
\V{d}_{j} = \sum_{i=1}^N \Gamma_{ji} \nabla x_{i}
+ \frac{(\VolumeFraction_j - w_j)}{n k_B T}  \nabla P + \V{d}_j^E ,
\label{eq:mass_flux_1}
\end{equation}
\modified{where
$\VolumeFraction_j$ are the volume fractions (defined after Eq. (\ref{EOS_Eqn})),
$P$ is the pressure,
$k_B$ is Boltzmann's constant,
$\V{\Gamma}$ is the matrix of activity coefficients (note that this matrix is the identity matrix for ideal mixtures, $\Gamma_{ij}=\delta_{ij}$).
The pressure (i.e., equation of state) and activities (i.e., chemical potentials) include all the contributions from short-ranged molecular interactions while $\V{d}_j^E$ gives the
diffusion driving force due to the long-range electrostatic interactions (details given below).}

The stochastic fluxes, which are derived from fluctuation dissipation, are given by,
\begin{equation}
\widetilde{\SpeciesFlux}_k = -\sqrt{2 k_B} \sum_{j=1}^{N} B_{kj} \V{\mathcal{Z}}_{j},
\label{eq:mass flux 2}
\end{equation}
where $\V{\mathcal{Z}}(\V r, t)$ denotes a collection of $N$ spatio-temporal white noise
random fields, i.e., a random Gaussian field with uncorrelated components,
\begin{equation}
\langle \V{\mathcal{Z}}_{k;\alpha}(\mathbf{r},t)
\V{\mathcal{Z}}_{k';\alpha'}(\mathbf{r}',t') \rangle =
\delta_{k,k'} \delta_{\alpha,\alpha'} \delta(\mathbf{r}-\mathbf{r}') \delta(t-t'),
\end{equation}
where $\alpha,\alpha'$ refer to $x,y,z$ components.
Finally, $\V{B}$ is a ``square root'' of the Onsager matrix,
$(\rho \bar{m}/k_B) \V{W} \V{\DiffusionMatrix} \V{W}$,
where $\V{W}$ is a matrix with elements of $\V{w}$ on the diagonal.
The matrix $\V{B}$ can be computed using a ``square root'' $\V{\DiffusionMatrix}^{\frac{1}{2}}$ of $\V{\DiffusionMatrix}$
(computed via a Cholesky or eigenvalue factorization), namely,
\begin{equation} \label{eq_B_def}
\M{B} = \sqrt{\frac{\rho\bar{m}}{k_B}} \V{W} \V{\DiffusionMatrix}^{\frac{1}{2}},
\end{equation}
so that $\V{B}\V{B}^T = (\rho \bar{m}/k_B) \V{W} \V{\DiffusionMatrix} \V{W}$.

The system is treated in the electroquasistatic
approximation~\cite{NumericalElectrodynamicsBook,grodzinsky2011field},
that is, the magnetic field is assumed constant
so Faraday's law for the electric field, $\mathbf{E}$, is
$\nabla \times \mathbf{E} = 0$.
The fluid mixture has permittivity $\epsilon = \epsilon_\mathrm{r} \epsilon_0$
where $\epsilon_\mathrm{r}$ is the
relative permittivity (also called the dielectric constant)
and $\epsilon_0$ is the vacuum permittivity.
By Gauss's law,
\begin{equation}
\nabla\cdot (\epsilon \V{E}) = -\nabla\cdot (\epsilon \nabla \ElectricPotential)
= q^\mathrm{f},
\label{eq:elec_field}
\end{equation}
where $\ElectricPotential$ is the scalar electric potential and
$\V{E} = - \nabla \ElectricPotential$.
We assume that the variation of permittivity is negligible and
take $\epsilon$ to be constant so $\epsilon \nabla^2 \ElectricPotential = -q^\mathrm{f}$.

For charged species, there is an additional contribution of the electric field to the diffusion
driving force \cite{DM_63, NernstPlanckModel}, leading to an additional
contribution to the deterministic flux,
\begin{equation}
\V{d}_j^E =
\frac{\bar{m}w_j}{k_B T} \left(\SpecificCharge_j -
\sum_{i=1}^N w_{i} \SpecificCharge_{i} \right) \nabla \ElectricPotential.
\end{equation}
Note that, when evaluating $\overline{\SpeciesFlux}_k $, we can simply use the expression
\begin{equation}
\V{d}_j^E =
\frac{\bar{m} w_j \SpecificCharge_j}{k_B T}  \nabla \ElectricPotential,
\end{equation}
since the vector $\V{w}$ is in the null space of $\V{\DiffusionMatrix}$ \cite{LowMachMultispecies}.

In the isothermal low Mach number model used in this paper, we neglect the barodiffusion
and thermodiffusion terms.
\modified{With these approximations, we can express the vector of species fluxes as,}
\begin{equation}
\SpeciesFlux = -\rho \V{W} \V{\DiffusionMatrix} \left( \V{\Gamma} \nabla\V{x} + \frac{\bar{m}\V{W}\V{z}}{k_B T}\nabla\ElectricPotential \right) - \sqrt{2k_B}\V{B}\V{\mathcal{Z}}.
\label{SpeciesTransportEqn_short}
\end{equation}
\new{In the \new{low dilution} limit where charge species are in trace quantities 
with one solvent species for which $x_N \rightarrow 1$,
we can omit the solvent from consideration and assume that $\V{\Gamma}$ is the identity matrix (ideal solution).
In this limit it can be shown that the $(N-1)\times(N-1)$ sub-block of $\M{\chi}$, corresponding to the
solutes, is approximately a diagonal matrix with entries
\begin{equation} \label{dilute_chi_kk}
\chi_{kk}=\frac{m_{k}D_{k}}{\bar{m}w_{k}},\quad k=1,\dots,N-1,
\end{equation}
where $D_k$ is the tracer diffusion coefficient of species $k$ in the solvent.
If we neglect gradients of pressure and temperature, for dilute solutions we recover the Nernst-Planck} model~\cite{newman2012electrochemical,NernstPlanckModel} for $k=1,..,N-1$,
 \begin{equation}\label{NernstPlanckEqn}
 \overline{\SpeciesFlux}_k =
 - D_k  \nabla \rho_k + \mathcal{M}_k \rho_k \V{E},
 \end{equation}
 where
 \begin{equation}\label{EinsteinRelationEqn}
 \mathcal{M}_k = \frac{D_k m_k \SpecificCharge_k}{k_B T}
 \end{equation}
 is the electrical mobility.
 Since $m_k \SpecificCharge_k$ is the molecular charge,
 (\ref{EinsteinRelationEqn}) is the Einstein relation.
 \fixed{Note that, as discussed at length in \cite{NernstPlanckModel}, the Nernst-Planck model is inconsistent
 with mass conservation and one should retain the solvent in the description as well, except, perhaps, for
 very dilute solutions.}

\subsubsection{Electroneutral Approximation}\label{sec:Electroneutral}

The electrostatic interactions between ions in the solution are screened by clouds of opposite charges. The length scale associated with this screening is the Debye length $\lambda_D$; at length scales much larger than $\lambda_D$, the fluid is neutrally charged. For dilute strong electrolytes Debye-H\"uckel theory ~\cite{dill2011molecular,grodzinsky2011field} gives,
\begin{equation}\label{DebyeLengthEqn}
\lambda_D = \left(\frac{\epsilon k_B T}{\sum_{k=1}^N \rho w_k m_k \SpecificCharge_k^2} \right)^{1/2}.
\end{equation}
The typical value of the Debye length in electrolytes is on the order of nanometers and in this work we aim to describe electrolytes down to microscopic scales below $\lambda_D$, where fluctuations are important. It should be noted, however, that when the Debye length is comparable to the molecular scales, the (fluctuating) continuum model used here may be inappropriate as complex chemical effects such as solvation layers may be important. It may be possible to ameliorate this problem by adjusting the activity and transport coefficients appropriately, but this is likely problem-specific.

At length scales larger than the Debye length, the diffusive motions of the ions are strongly coupled by the affinity to maintain electroneutrality \cite{Electrophysiology_Peskin_Review}, \modified{$q^\mathrm{f}=\rho \sum_{k=1}^N w_k \SpecificCharge_k=0$}.
\fixed{To a rough approximation, in the electroneutral limit one obtains for the species densities a} Fick's law with effective diffusion coefficients, known as the Nernst-Hartley diffusion coefficients~\cite{kontturi2008ionic,cussler2009diffusion,grodzinsky2011field}.
Note that this is essentially equivalent to ambipolar diffusion in plasmas,
which results from the quasineutrality approximation for ion and electron transport.
\fixed{A more careful mathematical derivation that shows that this common effective diffusion approximation is incomplete can be found in Section 5.2 in the review \cite{Electrophysiology_Peskin_Review}.}
Adding fluctuations to the electroneutral approximation is {\em not}
as trivial as simply adding the same unmodified stochastic fluxes $\widetilde{\SpeciesFlux}_k$ given in (\ref{eq:mass flux 2});
the stochastic fluxes must also be projected onto the charge-neutrality constraint.
\fixed{In particular, the fluctuating electroneutral equations ought to be} in detailed balance with respect to a Gibbs-Boltzmann distribution with free energy (dominated by entropy of mixing for ideal solutions) {\em constrained} by $w_k \SpecificCharge_k=0$, rather than the original free energy of the solution without the neutrality constraint. In this work we consider the full dynamics rather than the electroneutral approximation, and we will discuss the electroneutral limit in more detail in future work.

\subsection{Momentum Transport}

The momentum transport equation  has the general form,
\begin{equation}\label{MomentumEqn}
\frac{\partial(\rho \V{v})}{\partial t} + \nabla\cdot (\rho \V{v} \V{v}^T) =
-\nabla P + \nabla\cdot \ViscousTensor + \nabla\cdot\MaxwellTensor + \rho\V{g},
\end{equation}
where $\ViscousTensor$ is the viscous stress tensor,
$\MaxwellTensor$ is the Maxwell stress tensor,
and $\V{g}$ is the gravitational vector.
Similar to the species fluxes,
the viscous stress tensor 
has deterministic and stochastic contributions,
$\ViscousTensor = \overline{\ViscousTensor} + \widetilde\ViscousTensor$.
Since the Maxwell stress tensor expresses reversible work, there is no stochastic contribution to $\MaxwellTensor$.

Assuming that the viscous stress tensor is unaffected by the electric field
then both $\overline{\ViscousTensor}$ and
$\widetilde{\ViscousTensor}$ are the same as for neutral fluids.
We use the formulation as given in
\cite{MultispeciesCompressible,FluctHydroNonEq_Book,Landau:Fluid} and
ignore bulk viscosity effects, so
$\overline{\ViscousTensor} = \eta\bar\nabla\V{v} \equiv \eta[\nabla\V{v} + (\nabla\V{v})^T]$,
with viscosity $\eta$.
The stochastic contribution to the viscous stress tensor is formally modeled as,
\begin{equation}
\widetilde\ViscousTensor = \sqrt{\eta k_B T}(\V{\mathcal{W}} + \V{\mathcal{W}}^T),
\end{equation}
where $\V{\mathcal{W}}(\V r,t)$ is a standard white noise Gaussian tensor
with uncorrelated components,
\begin{equation}
\langle \V{\mathcal{W}}_{k;\alpha}(\mathbf{r},t)
\V{\mathcal{W}}_{k';\alpha'}(\mathbf{r}',t') \rangle =
\delta_{k,k'} \delta_{\alpha,\alpha'} \delta(\mathbf{r}-\mathbf{r}') \delta(t-t').
\end{equation}

In the absence of a magnetic field \cite{Landau1984electrodynamics},
\begin{equation}
\MaxwellTensor_{ij} = \epsilon E_i E_j - \half \epsilon E^2 \delta_{ij}.
\end{equation}
For a dielectric fluid with constant permittivity,
$\epsilon \nabla\cdot \V{E} = q^\mathrm{f}$, so that
the resulting force density on the fluid is,
\begin{equation}
\V{f}^\mathrm{E} = \nabla\cdot \MaxwellTensor = q^\mathrm{f} \V{E}
= - q^\mathrm{f} \nabla \ElectricPotential,
\end{equation}
which is simply the Lorentz force.




\subsection{Low Mach Number Model}

In this paper we will focus on systems with ions dissolved in a
neutral liquid solvent.
For these systems, the characteristic fluid velocity is small
compared to the sound speed (i.e., the Mach number Ma = $U/c \lesssim 0.1$)
and sound waves do not significantly
affect the thermodynamics of the system.
\modified{In this setting}, we can use a low Mach number approximation, \modified{which} can be derived from the fully
compressible equations by performing an asymptotic
analysis in Mach
number \cite{IncompressibleLimit_Majda,ZeroMachCombustion}.
The low Mach number model removes acoustic wave propagation from the system, resulting in a
system that can be efficiently integrated over advective time scales.

In our isothermal low Mach number model \cite{LowMachExplicit,LowMachMultispecies},
the momentum equation is recast as,
\begin{equation}\label{LowMachMomEqn}
\frac{\partial(\rho\V{v})}{\partial t}+ \grad\cdot\left(\rho\V{v}\V{v}^T\right) =
-\nabla\pi + \nabla\cdot \ViscousTensor  + \nabla\cdot\MaxwellTensor + \rho\V{g},
\end{equation}
\modified{where $\pi$ is a perturbational pressure} and the mass density equations (\ref{SpeciesTransportEqn}) and
(\ref{SpeciesTransportEqn_short}) are unchanged.
In order to mathematically close the evolution equations momentum and mass densities,
we require an additional relationship between the variables.  Typically this
is accomplished by supplying an equation of state; here we instead
require that the total mass density is a specified function of the
local composition.
In particular, we consider mixtures of incompressible fluids that do not change
volume upon mixing.
This leads to a constraint of the form \cite{LowMachMultispecies}
\begin{equation}\label{EOS_Eqn}
\sum_{k=1}^{N}\frac{\rho_{k}}{\bar{\rho}_{k}} = \rho\sum_{k=1}^{N}\frac{w_{k}}{\bar{\rho}_{k}}=1
\quad \rightarrow \quad
\rho(\V{w}) = \left(\sum_{k=1}^{N}\frac{w_{k}}{\bar{\rho}_{k}}\right)^{-1},
\end{equation}
where $\bar{\rho}_k$ is the (potentially hypothetical) pure-component density of species $k$.
This equation plays the role of the equation of state and gives
volume fractions $\VolumeFraction_k = \rho_{k}/\bar{\rho}_{k}$.
\modified{As detailed in Section \ref{ExamplesSection}, (\ref{EOS_Eqn}) can accurately approximate {\em any} mixture,
at least over a limited range of concentrations, with the appropriate 
choice for the constants $\bar{\rho}_{k}$.}
We can recast (\ref{EOS_Eqn}) into a divergence constraint on the velocity
field.  Taking the Lagrangian derivative of $\rho$,
\begin{equation}
\frac{D\rho}{Dt} = \sum_{k=1}^N \frac{\partial\rho}{\partial w_k}\frac{Dw_k}{Dt},
\end{equation}
and substituting in the density equation (\ref{ContinuityEqn}),
species equation (\ref{SpeciesTransportEqn}),
and derivatives of $\rho$ with respect to $w_k$ defined by (\ref{EOS_Eqn}),
we arrive at the velocity constraint \cite{LowMachMultispecies},
\begin{equation}\label{VelocityConstraintEqn}
\nabla \cdot \V{v} = - \nabla \cdot \left(
\sum_{k=1}^{N} \frac{\SpeciesFlux_k}{\bar{\rho}_k} \right)
\equiv -\nabla\cdot(\SpeciesFlux^T\V{\bar\nu}).
\end{equation}
where $\V{\bar\nu} = (\bar{\rho}_1^{-1},\bar{\rho}_2^{-1}, ..., \bar{\rho}_N^{-1})$
\modified{is the vector of species specific volume}.

Note that if the species fluxes vanish (e.g., immiscible mixtures) or the species are mechanically equivalent
(i.e., $\bar{\rho}_i = \bar{\rho}_j$ for all $i,j$) the model recovers the
familiar incompressibility constraint $\nabla \cdot \V{v} = 0$.
As discussed in \cite{LowMachExplicit},
the constraint on the velocity field ensures that the densities remain on the
equation of state (\ref{EOS_Eqn}).

We would like to point out that the inclusion of an energy evolution equation,
as well as the incorporation of a generalized equation of state, is a subject
for future work.  The model would be similar to other low Mach number
models with energy evolution \cite{ABRZ:I, LaminarFlowChemistry}
except that we would need to include an Ohmic
heating term due to the motion of charges in the presence of an electric field.

Altogether, our model equations consist of density transport
(\ref{SpeciesTransportEqn}), with mass fluxes (\ref{SpeciesTransportEqn_short}),
and momentum evolution (\ref{LowMachMomEqn}),
all constrained by the equation of state (\ref{VelocityConstraintEqn}).

\section{Structure Factors}\label{sec:Structure}

Some of the key measurements we use to validate
our numerical methodology involve the structure factor in both equilibrium
and non-equilibrium systems.\
\new{These results will also elucidate the role of the Debye length and the relation
between fluctuating hydrodynamics and the classical Debye-Huckel theory.}

\new{
We will need some matrix notation and relationships in our derivation.
The Jacobian of the transformation from mass to mole fractions is given by,
\begin{equation}
\frac{\partial\V{x}}{\partial\V{w}} = \left(\V{X} - \V{x}\V{x}^T\right)\V{W}^{-1},
\end{equation}
where capital $\V{W}$ and $\V{X}$ are diagonal matrices with elements $\V{w}$ and $\V{x}$.
Using this the diffusive flux can be recast in terms of gradient of mass fractions,
\begin{equation}\label{F_grad_w}
\rho\V{W}\V{\chi}\V{\Gamma}\nabla\V{x} =
\rho\V{W}\V{\chi}\V{\Gamma}(\V{X}-\V{x}\V{x}^T)\V{W}^{-1}\nabla\V{w}.
\end{equation}
For dilute solutions, we can eliminate the solvent from consideration and use the same equations, approximating 
$\V{X}-\V{x}\V{x}^T$ with $\V{X}=\bar{m} \M{M}^{-1} \M{W}$, and expressing
\begin{equation}\label{eq:low_dilution}
\M W\M{\chi}\M W \approx \bar{m}^{-1}\M M\M D\M W,
\end{equation}
where we have used (\ref{dilute_chi_kk}). Here $\V{M}$ is a diagonal matrix containing the molecular masses on the diagonal, and $\M{D}$ is a diagonal matrix containing the tracer diffusion coefficients on the diagonal.
}

\subsection{Equilibrium Fluctuations} \label{StructureFactor}

A key quantity in the stochastic setting is the
spectrum of fluctuations at equilibrium, specifically, the
static (equal time) structure factor for mass fractions,
\begin{equation}\label{StructureFactorDefinitionEqn}
S_w^{ij}(\V{k}) = \left\langle
\left( \widehat{\delta w}_i(\V{k},t) \right)
\left( \widehat{\delta w}_j(\V{k},t) \right)^*
\right\rangle,
\end{equation}
where $\delta\V{w} = \V{w} - \V{w}_\mathrm{eq}$ is the fluctuation
about the equilibrium state $\V{w}_\mathrm{eq}$, hat denotes Fourier transform,
star denotes complex conjugation,
and $\V{k}$ is the wavevector.
In the absence of charged species, $S_w^{ij}(\V{k})$ is
independent of $\V{k}$ \cite{LowMachMultispecies}, but in an electrolyte this is not the case.

To obtain an expression for $\V{S}_w$ we linearize (\ref{SpeciesTransportEqn}) and (\ref{eq:elec_field})
about an equilibrium state that is charge neutral and has zero velocity.
For this analysis, the perturbational quantities are denoted with a $\delta$.  The remaining
quantities refer to the constant value at equilibrium.
For the electric field we have
\begin{eqnarray}
- \epsilon \nabla^2 (\Phi + \delta \Phi)
&=& (\rho + \delta \rho) \V{z}^T ( \V{w} + \delta \V{w}) \nonumber\\
&=& \rho\V{z}^T\V{w} + (\delta \rho) \V{z}^T \V{w} + \rho \V{z}^T  \delta \V{w} + \mathcal{O}(\delta^2).
\end{eqnarray}
Charge neutrality gives $ \Phi = 0$ and $\V{z}^T \V{w} = 0$ so that
to leading order
\begin{equation}
- \epsilon \nabla^2  \delta \Phi =  \rho \V{z}^T  \delta \V{w}.
\end{equation}
If we now linearize (\ref{SpeciesTransportEqn}) we obtain
\begin{equation}
\rho \partial_t \delta \V{w} = -\rho \V{W} \V{\DiffusionMatrix} \left( \V{\Gamma} \nabla^2 \delta \V{x}
 +\frac{\bar{m}\V{W} \V{z}}{k_B T}  \nabla^2 \delta \ElectricPotential \right) - \nabla \cdot \widetilde{\SpeciesFlux}.
\end{equation}
Expressing the flux in terms of mass fractions as in (\ref{F_grad_w}),
using (\ref{eq_B_def}), and combining with the equation of the perturbational electric potential, we obtain
\begin{equation}
\rho \partial_t \delta \V{w} =
-\rho \V{W} \V{\DiffusionMatrix} \left( \V{\Gamma} \left(\V{X} - \V{x}\V{x}^T\right
)\V{W}^{-1} \nabla^2 \delta \V{w} -
\frac{\bar{m}\rho\V{W} \V{z}\V{z}^T}{k_B T {\epsilon}} \delta \V{w} \right) -
\nabla \cdot \left( \sqrt{2 k_B} \V{B} \V{\mathcal{Z}} \right).
\end{equation}
Taking the Fourier transform, we have
\begin{eqnarray}
\partial_t \widehat{\delta \V{w}} (\V{k})
&=& {\V{W}} \V{\DiffusionMatrix} \left( k^2 \V{\Gamma} (\V{X}-\V{x} \V{x}^T) {\V{W}}^{-1}
+ \frac{\bar{m}\rho\V{W} \V{z} \V{z}^T}{k_B T \epsilon}
\right) \widehat{\delta \V{w}}(\V{k})
- i \sqrt{\frac{2}{n}} \V{k}^T \V{W} \V{\DiffusionMatrix}^{\frac{1}{2}} \V{\widehat{\mathcal{Z}}}  \label{M_OU}\\
&=& \V{\mathcal{M}} \widehat{\delta \V{w}}(\V{k}) + \V{\mathcal{N}} \V{\widehat{\mathcal{Z}}}  \label{M_OU2},
\end{eqnarray}
where $\V{\mathcal{M}}$ and $\V{\mathcal{N}}$ are two constant matrices.
This is the equation for a multivariate Ornstein-Uhlenbeck process with stationary covariance that satisfies~\cite{GardinerBook}
\begin{equation}\label{OUeqn}
\V{\mathcal{M}} \V{S}_w + \V{S}_w \V{\mathcal{M}}^* = -\V{\mathcal{N}}\V{\mathcal{N}}^*,
\end{equation}
supplemented by the constraint that $\V{S}_w$ is symmetric, and that the row and column
sums of $\V{S}_w$ are zero because the mass fractions sum to one.
\deleted{; alternatively,
and more easily in this case, one can remove the solvent species from consideration and
only consider fluctuations in the concentrations of the dilute species.}


It can be shown that the solution to (\ref{OUeqn}) for a system in thermodynamic equilibrium is a simple rank-1 correction of
the static structure factor without the charges,
\begin{equation} \label{S_k_eq_general}
\M{S}_w = \M{S}_0 - \frac{1}{\left( k^2\lambda_D^{2}+1 \right)} \frac{\M{S}_0 \V{z} \V{z}^T \M{S}_0}{\V{z}^T \M{S}_0 \V{z}},
\end{equation}
where \modified{a generalized} Debye length can be written in matrix form as
\begin{equation} \label{Debye_general}
\lambda_D^{-2}=\frac{\rho^2}{\epsilon k_{B}T}\V z^{T} \M{S}_0 \V z.
\end{equation}
Here the structure factor for a mixture of uncharged species (i.e., for $\V{z} = 0$) is \cite{LowMachMultispecies}
\begin{equation}
\M{S}_0 = \lim_{k \lambda_D\rightarrow\infty} \M{S}_w(\V{k})
= \frac{\bar{m}}{\rho}\left(\M W-\V w\V w^{T}\right)\left[\M{\Gamma}\left(\M X-\V x\V x^{T}\right)+\V 1\V 1^{T}\right]^{-1}\left(\M W-\V w\V w^{T}\right),
\end{equation}
where $\V{1}$ is the vector of 1's.  Note that
\begin{equation}
\lim_{k\lambda_D\rightarrow 0} \M{S}_w(\V{k})\V{z} = 0
\end{equation}
as expected in the limit of electroneutrality.
Also note that the equilibrium static structure factor is a purely thermodynamic quantity that is independent of the dynamics, notably,
it is independent of the diffusion matrix. It can therefore also be derived from a free energy argument, in which an electrostatic contribution to the free energy is combined with the entropy of mixing (not shown in this paper).

For an electrolyte solution that is close to ideal, the explicit formula for $\M{S}_0$ is simpler (c.f. (D3) in \cite{LowMachMultispecies}),
and substituting this in (\ref{S_k_eq_general}) gives the equilibrium structure factor for a charged ideal mixture,
\begin{equation} \label{S_w_ideal}
\M{S}_w = \rho^{-1} \left( \M{I}-\V{w}\V{1}^T \right) \left[
\M{W}\M{M} - \frac{1}{\left( k^2\lambda_D^{2}+1 \right)} \frac{\M{W}\M{M} \V{z} \V{z}^T \M{M} \M{W}}{\V{z}^{T}\left(\M M\M W\right)\V{z}}
\right] \left( \M{I}-\V{1}\V{w}^T \right),
\end{equation}
where $\V{M}$ is a diagonal matrix containing the molecular masses $\V{m}$ on the diagonal.
If one is interested only in the solvent species in a dilute solution, the structure factor for the solvent species is given by the above formula without the projectors $\left( \M{I}-\V{w}\V{1}^T \right)$ and $\left( \M{I}-\V{1}\V{w}^T \right)$;
for a binary solution the resulting structure factor is in agreement with Berne and Pecora \cite{berne_pecora}.
For an ideal solution the Debye length is given by (\ref{DebyeLengthEqn}), which can be written in matrix notation as,
\begin{equation} \label{lambda_D_matrix}
\lambda_D^{-2} = \frac{\rho}{\epsilon k_{B}T}\V z^{T}\left(\M M\M W\right)\V z.
\end{equation}
It is significant that (\ref{Debye_general}) allows one to generalize the definition of the Debye length to
non-ideal electrolyte mixtures.

\modified{In the context of electrolytes, the specific charge $\bar{z}=\V{z}^T \V{w}$ is an important scalar quantity whose structure factor $S_{\bar{z}}$ is related to $S_w$ by
\begin{equation}
S_{\bar{z}}(\V{k}) = \left\langle
\left( \V{z}^T\widehat{\delta \V{w}}(\V{k},t) \right)
\left( \V{z}^T\widehat{\delta \V{w}}(\V{k},t) \right)^*
\right\rangle = \V{z}^T S_w \V{z}.
\end{equation}
Using the generalized definition of the Debye length \eqref{Debye_general} allows to conveniently express it as:
\begin{equation}\label{S_z}
S_{\bar{z}}(\V{k}) = \left( \V{z}^T S_0 \V{z} \right) \frac{k^2}{\lambda_D^{-2}+k^2} = \frac{\epsilon k_B T}{\rho^2}\frac{k^2}{1+k^2 \lambda_D^2}
\end{equation}
The fact that $S_{\bar{z}}(\V{k})$ tends to zero for small wavenumbers is a manifestation of the transition to the electroneutral regime at large length scales.
}

\subsection{Relation to Debye-Huckel theory}

\modified{In this section we relate the results derived in Section \ref{StructureFactor} to Debye-Huckel (DH) theory relying heavily on the excellent review article by Varela \textit{et al.}~\cite{Electrolytes_DH_review}. It has been known for some time that DH theory can be related to fluctuating field theories that include long-ranged Coulomb interactions, see Section 4 in \cite{Electrolytes_DH_review} for a review. In the Gaussian approximation to the fluctuations one recovers the classical Debye-Huckel theory, showing that it accounts for the corrections to the thermodynamic properties of the electrolyte mixture due to the charge fluctuations occurring at length scales below the Debye length. In this section we show the connection between classical DH theory and linearized fluctuating hydrodynamics \footnote{We believe that nonlinear corrections predicted by nonlinear field theories are also consistently captured by nonlinear fluctuating hydrodynamics, but this merits further study.} by deriving two key results of DH theory. Only a few simple steps are required, demonstrating the analytical power of the fluctuating hydrodynamic approach.

One of the key predictions of DH theory, leading to the introduction of the concept of screening and the Debye length, is the DH formula for the pair correlation function between solvent species $i$ and $j$,
\begin{equation}\label{DH_g_r}
g_{ij} = 1+h_{ij} = 1 - \frac{q_i q_j}{4 \pi \epsilon k_B T } \cdot \frac{1}{r} \exp \left(-\frac{r}{\lambda_D}\right),
\end{equation}
where $q_k=m_k z_k$ is the molecular charge.
We now show that it is relatively straightforward to obtain this result from the structure factor (\ref{S_w_ideal}) obtained by fluctuating hydrodynamics. 

First, in order to be consistent with the classical derivation \cite{Electrolytes_DH_review} we assume that the solution is ideal and eliminate the solvent species from consideration, giving the solute structure factor
\begin{equation}
\widetilde{\M{S}}_w  = \rho^{-1} \M{W}\M{M} -
\frac{1}{\epsilon k_B T \left( k^2 + \lambda_D^{-2} \right)} \M{W} \V{q} \V{q}^T \M{W}.
\end{equation}
We can convert this into the structure factor for mole fractions used in \cite{Electrolytes_DH_review} by noting that $\V{x}=\bar{m}\M{M}^{-1}\V{w}$,
\begin{equation}
\widetilde{\M{S}}_x =
\left\langle \left( \widehat{\delta \V x} \right) \left( \widehat{\delta \V x} \right)^* \right\rangle
= \bar{m}^2 \M{M}^{-1} \widetilde{\M{S}}_w \M{M}^{-1},
\end{equation}
to obtain (compare to Eq. (180) in \cite{Electrolytes_DH_review})
\begin{equation}
\widetilde{S}_x^{ij} = n^{-1}  x_i \delta_{ij} -
\frac{1}{\epsilon k_B T \left( k^2 + \lambda_D^{-2} \right)} x_i x_j q_i q_j.
\end{equation}
The pair correlation function is related to the structure factor via the formula (see Eq. (179) in \cite{Electrolytes_DH_review})
\begin{equation}\label{DH_h_ij}
\hat{h}_{ij}(k) = \frac{\widetilde{S}_x^{ij} - n^{-1}  x_i \delta_{ij}}{x_i x_j} =- \frac{1}{\epsilon k_B T \left( k^2 + \lambda_D^{-2} \right)} q_i q_j.
\end{equation}
The DH equation (\ref{DH_g_r}) now follows from a simple conversion of $\hat{h}(k)$ from Fourier space to real space, demonstrating that (\ref{S_w_ideal}) is consistent with the standard DH theory.

Another key result of DH theory is that the change in (renormalization of) the internal energy density due to electrostatic interactions is
\begin{equation}\label{DH_U_e}
u_e = -\frac{k_B T}{8\pi \lambda_D^3} \sim \frac{1}{\sqrt{T}}.
\end{equation}
From this relation, one can obtain the corrections to all other thermodynamic quantities such as the Gibbs free energy density and the osmotic pressure contribution to the equation of state \cite{Electrolytes_DH_review}. Here we show how to obtain this relation from (\ref{S_z}), thereby demonstrating that it generalizes beyond just ideal solutions. 

In DH theory one obtains this relationship by integrating the pair correlation function (\ref{DH_g_r}) times the Coulomb potential. This calculation is actually simpler and more transparent in Fourier space. In real space, the electrostatic contribution to the internal energy density is
\begin{equation}
u_e = \frac{\rho}{2 V} \int \left\langle \left( \V{z}^T \delta\V{w}(\V{r}) \right) \delta\phi(\V{r}) \right\rangle \; d\V{r}.
\end{equation}
Recalling the Poisson equation relating $\delta\phi$ with $\delta\V{w}$ and using Parseval's formula to convert this into an integral in Fourier space, we obtain
\begin{equation}
u_e = \frac{\rho^2}{2 (2\pi)^3 \epsilon} \int k^{-2} \V{z}^T
\left\langle   \left(\widehat{\delta \V{w}}\right) \left(\widehat{\delta \V{w}}\right)^T  \right\rangle \V{z} \; d\V{k}
= \frac{\rho^2}{(2\pi)^2 \epsilon } \int_0^\infty S_{\bar{z}}(k) dk.
\end{equation}
As written, the integral diverges, however, the same problem also appears in the real space derivation, as reviewed in \cite{Electrolytes_DH_review}. Because of global electroneutrality, the non-convergent part of the integral is actually zero, and one should only include the contribution to $S_{\bar{z}}$ (see (\ref{S_z})) that comes from the electrostatic interactions while excluding the part coming from equilibrium fluctuations in the absence of charges, just as we subtracted the equilibrium piece $n^{-1}  x_i \delta_{ij}$ from $\widetilde{S}_x^{ij}$ in (\ref{DH_h_ij}). This gives
\begin{equation}
u_e = \frac{\rho^2}{(2\pi)^2 \epsilon} \int_0^\infty \left(S_{\bar{z}} - \V{z}^T S_0 \V{z} \right) dk
 = - \frac{k_B T}{(2\pi)^2 \lambda_D^2} \int_0^\infty \frac{1}{1+k^2 \lambda_D^2} dk = -\frac{k_B T}{8\pi \lambda_D^3},
\end{equation}
in agreement with the DH theory expression (\ref{DH_U_e}). 

These results demonstrate that our fluctuating hydrodynamics formalism reproduces Debye-Huckel theory. Finally, note that the theory presented here is specifically for three dimensional systems. In two dimensions, the above integrals diverge logarithmically in the infinite system size limit due to the pathological logarithmic divergence of the Coulomb potential in two dimensions.
}

\subsection{Structure Factor for the Giant Fluctuations in Non-equilibrium systems}\label{StructFactGiantFluct}

In this section, we derive the theoretical values for the structure factors of the giant fluctuations
that develop in non-equilibrium systems, following similar calculations we performed in Refs.~\cite{MultispeciesCompressible, MultispeciesChemistry}.
It is known that a multispecies mixture subjected to concentrations gradients develops long-range correlations, and that the
structure factor of the fluctuations of the concentrations varies according to a power law $k^{-4}$ \cite{FluctHydroNonEq_Book}.
These giant fluctuations arise due to the advection of the concentration fluctuations by the random velocity field,
and therefore, these simulations require
the complete hydrodynamic solver including the fluctuating momentum equation.
We seek to examine how a system of charged species deviates from this law.
We assume that there is a macroscopic gradient
of mass fractions in the $y$-direction for all species,
\begin{equation}
g_k = \frac{\partial w_k}{\partial y},
\end{equation}
and we seek to determine the structure factor of the fluctuations with respect to the
wavenumber perpendicular to the gradient, $k_{\perp}=\sqrt{k_x^2+k_z^2}$. Without loss of generality we can set $k_z=0$ henceforth.

Linearizing (\ref{SpeciesTransportEqn}) about the non-equilibrium state results in
\begin{equation}
\rho \left( \partial_t \delta \V{w} + \V{g} \delta v_y \right) =
-\nabla \cdot \delta \bar{\V{\mathcal{F}}} - \nabla \cdot \tilde{\V{\mathcal{F}}}.
\label{LinearizedNonEq}
\end{equation}
Following Ref.~\cite{FluctHydroNonEq_Book},
we can obtain a system involving only $\delta v_y$
by applying a $\grad \times \grad \times $ operator to the momentum equation, leading to
\begin{equation}
\rho \partial_t (\grad^2 \delta v_y) = \eta \grad^2 (\grad^2 \delta v_y) + \grad \times \grad \times (\grad \cdot \widetilde\ViscousTensor_{\cdot;y})
\label{projectedMomentum}
\end{equation}
where $\widetilde\ViscousTensor_{\cdot;y}$ is the second column of the matrix $\widetilde\ViscousTensor$.


In the limit of large Schmidt number (overdamped or steady Stokes limit), we can neglect inertia and set the left hand side to 0.
The Fourier transform of \eqref{projectedMomentum} simply becomes, when $k_y=0$,
\begin{equation}
\widehat{\delta v_y} = i\sqrt{\frac{2k_B T}{\eta}}\frac{1}{k_x} \mathcal{V}(t),
\label{overdamped2}
\end{equation}
where $\sqrt{2}\mathcal{V}=\widehat{{\mathcal{W}}}_{x;y}+\widehat{{\mathcal{W}}}_{y;x}$ is a
white noise Gaussian process.
\deleted{Since ${\mathcal{W}}_{x;y}$ and ${\mathcal{W}}_{y;x}$ are uncorrelated white noise Gaussian random variables, their
sum can be replaced by a single random variable $\sqrt{2}\mathcal{V}$ where $\mathcal{V}(t)$ is a
white noise Gaussian process.}

Inserting Eq.~\eqref{overdamped2} into the Fourier transform of Eq.~\eqref{LinearizedNonEq} finally yields
\begin{equation}
\partial_t \widehat{\delta \V{w}} (k_x)
= \V{\mathcal{M}} \widehat{\delta \V{w}}(k_x) + \V{\mathcal{N}} \widehat{\V{\mathcal{Z}}}(t) + \V{\mathcal{N}}_\text{adv} \widehat{{\mathcal{V}}}(t),
\end{equation}
with
\begin{equation}
\V{\mathcal{N}}_\text{adv} = -i \sqrt{\frac{2k_B T}{\eta}} \frac{1}{k_x} \V{g},
\end{equation}
where $\V{\mathcal{M}}$ and $\V{\mathcal{N}}$ are given in Eqs.~\eqref{M_OU}-\eqref{M_OU2}.
Since $\mathcal{V}$ and the components of $\V{\mathcal{Z}}$ are uncorrelated,
the structure factor for the giant fluctuations is the solution to
\begin{equation}\label{OUeqnGiantFluct}
\V{\mathcal{M}} \V{S}_w + \V{S}_w \V{\mathcal{M}}^* = -\V{\mathcal{N}}\V{\mathcal{N}}^* -\V{\mathcal{N}}_\text{adv}\V{\mathcal{N}}^*_\text{adv}.
\end{equation}
In the remainder of this section we will focus on the nonequilibrium contribution $\V{S}_{\textrm{neq}}$ to the structure factor due to advection, obtained by solving
\begin{equation}\label{OUeqnGiantFluct_neq}
\V{\mathcal{M}} \V{S}_{\textrm{neq}} + \V{S}_{\textrm{neq}} \V{\mathcal{M}}^* = -\V{\mathcal{N}}_\text{adv}\V{\mathcal{N}}^*_\text{adv}.
\end{equation}

We have solved these equations for the case of a low-dilution solution of two charged species in a neutral solvent, using the
symbolic algebra software Maple. The general solution is analytically complex and we omit it here for brevity,
but note the following observations.
First, for scales much smaller than the Debye length, the charges have no effect and one recovers the
well-known $~k_x^{-4}$ spectrum for the giant fluctuations in a low-density solution of uncharged species \cite{FluctHydroNonEq_Book,LowMachMultispecies}:
\begin{equation}
\M{S}_{\textrm{neq}}(k_x\lambda \gg 1) = \M{S}_{(n)} = {\frac {k_{{B}}T}{{k}_x^{4}\eta}}
\left[ \begin {array}{cc} {\frac {{g_{{1}}}^{2}}{D_{{1}}}}&2\,{\frac {g_{{2}}g_{{1}}}{D_{{1}}+D_{{2}}}}\\ \noalign{\medskip}2\,
{\frac {g_{{2}}g_{{1}}}{D_{{1}}+D_{{2}}}}&{\frac {{g_{{2}}}^{2}}{D_{{2}}}}\end {array} \right]
\end{equation}
Note that the nonequilibrium concentration fluctuations in the different species are strongly correlated to each other
since they are both driven by the {\em same} velocity fluctuations \cite{DiffusionJSTAT}.
For scales much larger than the Debye length, one can use the electroneutral approximation and treat both ions as one species diffusing
with an effective ``ambipolar'' diffusion coefficient that is a weighted harmonic average of the self diffusion coefficients of the two ions \cite{cussler2009diffusion},
\begin{equation}
D_{\text{amb}}={\frac {D_{{1}}D_{{2}} \left( m_{{1}}z_{{1}}-
m_{{2}}z_{{2}} \right) }{D_{{1}}m_{{1}}z_{{1}}-D_{{2}}m_{{2}}z_{{2}}}},
\label{ambipolarCoeff}
\end{equation}
and use the well-known theory for a mixture of two uncharged liquids \cite{FluctHydroNonEq_Book}.
For scales comparable to the Debye length, the general result is tedious and we evaluate the complex analytical formulas numerically. In the next section, in Figure~\ref{GiantFluctStruct},
we show comparisons between the theoretical results and results obtained from the simulation method presented in Section~\ref{NumericsSection}.

If the two species have the same diffusion coefficient (even if they have different masses), we obtain that $\M{S}_{\textrm{neq}}=\M{S}_{(n)}$, that is,
the charges do {\em not} affect the giant fluctuations. When the diffusion coefficients are different, 
all of the components of the nonequilibrium structure factors still have the same power law divergence $k_x^{-4}$ at all wavenumbers, however, the coefficient in front of $k_x^{-4}$ changes for $k_x\lambda \ll 1$.
For example, for $D_2=r D_1$, and equal masses, $m_2=m_1$,
the ratio between the cross-correlation of the nonequilibrium fluctuations with and without charges is given by
\begin{equation} \label{Giant_S_k_simple}
\frac{{S}^{12}_{\textrm{neq}}}{{S}_{(n)}^{{12}}}={\frac {4\,{k}_x^{2}{\lambda}^{2}r+{r}^{2}+2\,r+1}
                              {4 \left( {k}_x^{2}{\lambda}^{2}+1 \right) r}}.
\end{equation}

\section{Numerical Methods}\label{NumericsSection}

The core numerical methodology is similar to our previous work for
neutral binary and multicomponent diffusive mixing 
\cite{LowMachExplicit,LowMachImplicit,LowMachMultispecies}.  
The overall numerical framework is a structured-grid
finite-volume approach
with cell-averaged densities and face-averaged (staggered) velocities.
We summarize the temporal discretization
below, and refer the reader to our previous works for details 
of the spatial discretization, noting that we choose standard
second-order stencils for derivatives and spatial averaging to satisfy
fluctuation-dissipation balance.
The main addition here is the electrostatic contribution to the mass fluxes and
the Lorentz force in the momentum equation.

Recall that our model equations consist of density transport 
(\ref{SpeciesTransportEqn}) and momentum evolution (\ref{LowMachMomEqn}) subject
to the constraint on the velocity field (\ref{VelocityConstraintEqn}).
The overall approach is
a second-order predictor-corrector for species densities and velocity,
developed in our prior work \cite{LowMachImplicit}.
The only change from the case of uncharged species is that computing the mass fluxes explicitly requires 
first solving a Poisson equation for the electric potential, which is a standard procedure done efficiently using a cell-centered multigrid solver.
Nevertheless, for the benefit of the reader, below we reproduce here a complete description of 
the time stepping algorithm used in the simulations reported here. 
We note that the algorithm used here is suitable for finite Reynolds number simulations 
\modified{which introduce a limitation on the time step size based on stability} restrictions. 
It is important to note that in \cite{LowMachImplicit} we also describe an {\em overdamped} algorithm 
in which we neglect the inertia of the fluid and solve a {\em steady} Stokes problem for the velocity 
instead of an unsteady one. That algorithm can also trivially be generalized to the charged case since the mass fluxes are computed explicitly in both algorithms.

In order to advance the velocities semi-implicitly subject to the
constraint on the velocity field, we have previously developed
a \modified{generalized} Stokes solver for this constrained evolution problem
(see \cite{LowMachImplicit, StokesKrylov}).
We advance the solution $(\V{v},\V{\rho})$ from $t^n$ to
$t^{n+1} = t^n + \Delta t$ using the following time-advancement scheme,
where the superscript on each term denotes its temporal location:
\begin{enumerate}

\item Obtain the electric potential by solving the Poisson equation,
\begin{equation}
\epsilon\nabla^2 \ElectricPotential^n = -(q^\mathrm{f})^n,
\end{equation}
and then compute the predictor mass fluxes
\begin{equation}
\SpeciesFlux^n = \left[-\rho \V{W} \V{\DiffusionMatrix} \left( \V{\Gamma} \nabla\V{x} + \frac{\bar{m}\V{W}\V{z}}{k_B T}\nabla\ElectricPotential \right)\right]^n - \sqrt{\frac{2k_B}{\Delta t\Delta V}}\V{B}^n\V{\mathcal{Z}}^{n:n+1},
\end{equation}
with cell volume $\Delta V$.
We use the notation $\V{\mathcal{Z}}^{n:n+1}$ to refer to the 
collection of random fields associated with this time step.
Note that this step is only needed when the algorithm is initialized.  Thereafter the mass fluxes and electric
potential have already been computed during Step 6 of the previous time step.

\item Update the species densities using a forward Euler predictor step,
\begin{equation}
\rho_k^{*,n+1} = \rho_k^n - \Delta t \nabla \cdot \SpeciesFlux_k^n - \Delta t\nabla\cdot(\rho_k\V{v})^n.
\end{equation}

\item Calculate corrector mass fluxes by first solving the Poisson equation
\begin{equation}
\epsilon\nabla^2 \ElectricPotential^{*,n+1} = -(q^\mathrm{f})^{*,n+1}.
\end{equation}
and then evaluating the fluxes explicitly,
\begin{equation}
\SpeciesFlux^{*,n+1} = \left[-\rho \V{W} \V{\DiffusionMatrix} \left( \V{\Gamma} \nabla\V{x} + \frac{\bar{m}\V{W}\V{z}}{k_B T}\nabla\ElectricPotential \right)\right]^{*,n+1} - \sqrt{\frac{2k_B}{\Delta t\Delta V}}\V{B}^{*,n+1}\V{\mathcal{Z}}^{n:n+1}.
\end{equation}

\item Compute a predicted velocity using a Crank-Nicolson discretization by solving \cite{StokesKrylov} the following Stokes system for velocity, $\V{v}^{*,n+1}$, and pressure, $\pi^{*,n+1}$,
\begin{eqnarray}
\frac{\rho^{*,n+1}\V{v}^{*,n+1} - \rho^n\V{v}^n}{\Delta t} + \nabla\pi^{*,n+1} &=& -\nabla\cdot(\rho\V{v}\V{v})^n + \rho^n\V{g}\nonumber\\
&&+ \half\nabla\cdot\left(\eta\bar\nabla\V{v}^n\right) + \half\nabla\cdot\left(\eta\bar\nabla\V{v}^{*,n+1}\right) \nonumber\\
&& + \nabla\cdot\sqrt{\frac{\eta k_B T}{\Delta t\Delta V}}(\V{\mathcal{W}}+\V{\mathcal{W}}^T)^{n:n+1} \nonumber\\
&&-\half(q^\mathrm{f}\nabla\ElectricPotential)^n - \half(q^\mathrm{f}\nabla\ElectricPotential)^{*,n+1},
\end{eqnarray}
\begin{equation}
\nabla\cdot\V{v}^{*,n+1} = -\nabla\cdot(\SpeciesFlux^T\V{\bar\nu})^{*,n+1}.
\end{equation}

\item Correct the species densities using a trapezoidal corrector,
\begin{equation}
\rho_k^{n+1} = \rho_k^n - \frac{\Delta t}{2} (\nabla \cdot \SpeciesFlux_k^n + \nabla \cdot \SpeciesFlux_k^{*,n+1}) - \frac{\Delta t}{2}\left[\nabla\cdot(\rho_k\V{v})^n + \nabla\cdot(\rho_k\V{v})^{*,n+1} \right].
\end{equation}

\item Solve the Poisson equation
\begin{equation}
\epsilon\nabla^2 \ElectricPotential^{n+1} = -(q^\mathrm{f})^{n+1},
\end{equation}
and calculate updated mass fluxes, noting we use a new set of stochastic fluxes,
formally associated with the {\em next} time step,
\begin{equation}
\SpeciesFlux^{n+1} = \left[-\rho \V{W} \V{\DiffusionMatrix} \left( \V{\Gamma} \nabla\V{x} + \frac{\bar{m}\V{W}\V{z}}{k_B T}\nabla\ElectricPotential \right)\right]^{n+1} - \sqrt{\frac{2k_B}{\Delta t\Delta V}}\V{B}^{n+1}\V{\mathcal{Z}}^{n+1:n+2}.
\end{equation}

\item Correct the velocity using a Crank-Nicolson discretization by solving the following Stokes system for velocity, $\V{v}^{n+1}$, and pressure, $\pi^{n+1}$,
\begin{eqnarray}
\frac{\rho^{n+1}\V{v}^{n+1} - \rho^n\V{v}^n}{\Delta t} + \nabla\pi^{n+1} &=& -\half\nabla\cdot(\rho\V{v}\V{v})^n - \half\nabla\cdot(\rho\V{v}\V{v})^{*,n+1} + \half(\rho^n+\rho^{n+1})\V{g}\nonumber\\
&&+ \half\nabla\cdot\left(\eta\bar\nabla\V{v}^n\right) + \half\nabla\cdot\left(\eta\bar\nabla\V{v}^{n+1}\right) \nonumber\\
&& + \nabla\cdot\sqrt{\frac{\eta k_B T}{\Delta t\Delta V}}(\V{\mathcal{W}}+\V{\mathcal{W}}^T)^{n:n+1}\nonumber\\
&&-\half(q^\mathrm{f}\nabla\ElectricPotential)^n - \half(q^\mathrm{f}\nabla\ElectricPotential)^{n+1},
\end{eqnarray}
\begin{equation}
\nabla\cdot\V{v}^{n+1} = -\nabla\cdot(\SpeciesFlux^T\V{\bar\nu})^{n+1}.
\end{equation}

\end{enumerate}

\subsection{Numerical Stability}
Since we treat the viscosity implicitly and all other terms explicitly, the largest stable computational time
step is dictated by one of three different effects;
the advective Courant-Friedrichs-Lewy (CFL) condition, the explicit
mass diffusion condition, and a stiffness associated with the electrostatic
driving force in the density equations.
Here we comment on the stability criteria related to each term.
Consider the mass density evolution equations,
\begin{equation}\label{eq:full mass density update}
\frac{\partial(\rho\V{w})}{\partial t} + \nabla\cdot(\rho\V{w}\V{v}) = 
\nabla\cdot\left(
\rho\V{W}\V{\chi}\V{\Gamma}\nabla\V{x} + 
\frac{\rho\bar{m}}{k_B T}\V{W}\V{\chi}\V{W}\V{z}\nabla{\ElectricPotential}
\right).
\end{equation}

The presence of the convective term requires the classical advective CFL time step
constraint,
\begin{equation}
\Delta t < \frac{\Delta x}{|v_{\rm max}|},
\end{equation}
where $v_{\rm max}$ is the largest magnitude velocity in the simulation.

Given (\ref{F_grad_w}), the explicit mass diffusion time step constraint is
\begin{equation}
\Delta t < \frac{\Delta x^2}{2d\beta_{\rm max}},
\end{equation}
where $d$ is the dimensionality of the problem, and $\beta_{\rm max}$ is the 
largest eigenvalue of $\V{W}\V{\chi}\V{\Gamma}(\V{X}-\V{x}\V{x}^T)\V{W}^{-1}$.
\new{
For dilute solutions, the diffusive flux can be simplified using (\ref{eq:low_dilution}) to
\begin{equation}
\rho\V{W}\V{\chi}\V{\Gamma}\nabla\V{x} \approx \rho\M{D}\nabla\V{w},
\end{equation}
to obtain the familiar stability restriction
\begin{equation} \label{dt_diff_dilute}
\Delta t < \frac{\Delta x^2}{2d \; \max_{1\leq k \leq N-1}{D_k}}.
\end{equation}
}

For the electrostatic driving force, note that
$\nabla^2\ElectricPotential = -\rho\V{w}^T\V{z}/\epsilon$. If we assume
that the prefactor multiplying the potential gradient in
(\ref{eq:full mass density update}) is roughly constant over
a small region, we can replace the divergence of the potential gradient with
the charge, and rewrite this term as
\begin{equation}
\nabla \cdot \left(\frac{\rho\bar{m}}{k_B T}\V{W}\V{\chi}\V{W}\V{z}\nabla{\ElectricPotential}\right)
\approx
-\frac{\rho\bar{m}}{\epsilon k_B T}\V{W}\V{\chi}\V{W}\V{z}(\V{z}^T\rho\V{w})
\end{equation}
If we consider the electric potential term in isolation, we can recast the equation
as a simple ODE, \modified{$d\V{w}/dt=-\V{\alpha}\V{w}$, where the matrix $\V{\alpha}$ is defined by
\begin{equation}
\V{\alpha} = \frac{\rho\bar{m}}{\epsilon k_B T}\V{W}\V{\chi}\V{W}\V{z}\V{z}^T.
\end{equation}
}
For our explicit temporal discretization, in order to avoid instability and negative densities,
we need a time step that satisfies the stability condition
$\Delta t < 1/{\alpha_{\rm max}}$,
where $\alpha_{\rm max}$ is the largest eigenvalue of $\V{\alpha}$.
\new{Since $\M{\alpha}$ is a rank-1 matrix, its only nonzero eigenvalue
corresponds to the eigenvector $\V{W}\V{\chi}\V{W}\V{z}$ and an eigenvalue
\begin{equation}
\alpha_{\rm max} = \frac{\rho\bar{m}}{\epsilon k_B T}\V{z}^T\V{W}\V{\chi}\V{W}\V{z}.
\end{equation}
For dilute solutions, we can use (\ref{eq:low_dilution}) to obtain
\begin{equation}
\alpha_{\rm max} = \frac{\rho}{\epsilon k_B T}\V{z}^T \M M\M D\M W \V{z}.
\end{equation}
If we assume all ions have the same diffusion coefficient, $\M D \approx D_0 \M I$,
and use (\ref{lambda_D_matrix}), we can express this in the physically intuitive form
$\alpha_{\rm max} = D_0 \lambda_D^{-2}$, giving an estimate for the stability restriction on the time step,
\begin{equation} \label{dt_electro_dilute}
\Delta t < \frac{\lambda_D^2}{\max_{1\leq k \leq N-1}{D_k}}.
\end{equation}
}

It is important to note that the electrostatic time step restriction is {\it not}
a function of grid spacing or the length scale of the problem, whereas the advective and mass
diffusion time steps scale with $\Delta x$ and $\Delta x^2$, respectively.
Thus, given the same fluid, if the length scales of the problem are sufficiently large, the time step will be dictated by the electrostatic driving force condition,
unless one makes use of the electroneutral approximation, which we will do in future work.
\new{Here we resolve the Debye length, $\Delta x < \lambda_D$, which implies that (\ref{dt_diff_dilute})
is more strict than (\ref{dt_electro_dilute}), justifying our explicit treatment of the electrostatic potential.}

\section{Numerical Examples}\label{ExamplesSection}

We now present some numerical examples that verify the accuracy of our approach.
Here we simulate salt (NaCl) dissolved in water at a
molarity comparable to seawater.  This model consists of three species; positively
charged Na, negatively charged Cl, and neutral water.
Parameters for this model are given in Table~\ref{ParametersTable}.
Here, $D_1$ and $D_2$ are the diffusion coefficients for sodium
and chloride ions in water \cite{LiGregory74} in the infinite dilution limit,
and $D_3$ is the self-diffusion
coefficient of water used in our neutral fluid study \cite{LowMachMultispecies}.
\begin{table}
  \centering
  \begin{tabular}{|c||c|c|c|}
     \hline
     Species & Sodium Ion & Chlorine Ion & Water \\
     \hline
     $m_k$ (g) &
     $3.82\times 10^{-23}$ & $5.89\times 10^{-23}$ & $3.35\times 10^{-23}$ \\
     $\SpecificCharge_k$ (C/g) &
     $4.2\times 10^{3}$ & $-2.72\times 10^{3}$ & 0 \\
     $\bar{\rho}_k$ (g/$\mathrm{cm}^3$) &
     3.17 & 3.17 & 1.0 \\
     $D_k$ ($\mathrm{cm}^2$/s) &
     $1.33\times 10^{-5}$ & $2.03\times 10^{-5}$ & $2.30\times 10^{-5}$ \\
     \hline
   \end{tabular}
  \caption{Fluid parameters in rationalized CGS units. The viscosity is
    $\eta = 1.05 \times 10^{-2}~\mathrm{g/cm~s}$
    for saltwater.
    The temperature is $300$~K,
    we assume ideal diffusion so the matrix of thermodynamic factors
    $\V{\Gamma}$ is the identity,
    and the relative permittivity is $\epsilon_\mathrm{r} = 78$.}
\label{ParametersTable}
\end{table}

We define two mixtures,
\begin{eqnarray}
\V{w}^{(1)} = (w_\mathrm{Na}^{(1)},w_\mathrm{Cl}^{(1)},w_{\mathrm{H}_2\mathrm{O}}^{(1)}) = (0.01088, 0.0168, 0.97232),\label{eq:w1}\\
\V{w}^{(2)} = (w_\mathrm{Na}^{(2)},w_\mathrm{Cl}^{(2)},w_{\mathrm{H}_2\mathrm{O}}^{(2)}) = (0.001088, 0.00168, 0.997232),\label{eq:w2}
\end{eqnarray}
and define our initial conditions for each test using these mixtures.
These two mixtures have Debye lengths of
$\lambda_D^{(1)} = 0.44$~nm and
$\lambda_D^{(2)} = 1.40$~nm.
The characteristic velocities in our examples are small enough that
each simulation is limited by either the electrostatic or mass diffusion time
step restriction.  For mixtures of fluids containing
(\ref{eq:w1}) and (\ref{eq:w2}),
the maximum allowable time step due to the
electrostatic stability condition is,
\begin{equation}
\Delta t < \frac{1}{\alpha_{\rm max}} = 1.16\times 10^{-10}~{\rm s}.\label{eq:electro_dt}
\end{equation}
The maximum allowable time step given by the mass diffusion stability 
condition is
\begin{equation}
\Delta t < \frac{\Delta x^2}{2d\beta_{\rm max}} = \frac{\Delta x^2}{2d(2.03\times 10^{-5} ~{\rm cm}^2.{\rm s}^{-1})}.\label{eq:massdiff_dt}
\end{equation}
Thus, in two dimensions ($d=2$), for
$\Delta x \gtrsim 10^{-7}$ cm, the time step is limited by the electrostatic driving force,
whereas for $\Delta x \lesssim 10^{-7}$ cm, the time step is limited by mass diffusion.

The procedure for determining the pure component densities used in our low
Mach velocity constraint (\ref{VelocityConstraintEqn}),
$\bar{\rho}_k$, from experimental data is
explained in \cite{LowMachMultispecies}. For saltwater we estimate
the values by taking the solutal expansion coefficient to be
40 cm$^3$/mol as given in Table I of \cite{MixedDiffusiveInstability}
and using the theory for density dependence on concentration for
dilute solutions used in \cite{LowMachMultispecies}.
The body force acceleration, $\V{g}$, is
zero in all of the following examples.

As in \cite{Diffusion_InfiniteDilution} we use
estimate the Maxwell-Stefan binary diffusion coefficients using the 
approximation,
\begin{equation} \label{MS_low_dilution}
D_{13} = D_1, \qquad D_{23} = D_2, \qquad D_{12} = \frac{D_1 D_2}{D_3}.
\end{equation}
The diffusion matrix $\V{\DiffusionMatrix}$ is computed
from $D_{ij}$ and $\V{w}$ using the iterative procedure presented in Appendix A
of \cite{LowMachMultispecies}.
Note that the validity of the assumptions used to derive (\ref{MS_low_dilution}) are questioned 
in \cite{MS_diffusion_ionic}, however, in the end, all that matters is that for a dilute solution,
ignoring the solvent species, \new{one obtains the familiar Fick's law for each of the solutes, without cross-diffusion.}

Note that the Schmidt number for this solution is $\textrm{Sc}=\eta/(\rho D_{k,{\rm max}}) \approx 500$,
which is quite large. Therefore, there will be some benefit in using the steady Stokes
approximation of the momentum equation and the associated overdamped algorithm described in \cite{LowMachMultispecies}.
Here we use relatively small time steps in order to control the error in the fluctuation spectrum
at large wavenumbers, and therefore continue to use the inertial formulation of the momentum equation.
Nevertheless, we can expect to see some (small) errors at the very largest wavenumbers since
the viscous Courant number is typically much larger than 1 (see right panel of Fig. 3 in \cite{LowMachMultispecies}).



\subsection{Deterministic Tests}\label{DeterministicTest}

To validate the implementation of the numerical method in a deterministic
setting, we diffuse a strip of saltwater into a less-salty ambient.
The two-dimensional domain is square with side length $L=3.6\times 10^{-5}$~cm
(a factor of $\sim 250$ larger than $\lambda_D^{(2)}$),
and periodic boundary conditions.
We initialize a horizontal strip in the center of the domain
with a width equal to $L/2$ to a saltier concentration, with
a smooth transition at each interface.
Specifically, we use
\begin{equation}
\V{w}(y) = \V{w}^{(2)} + \frac{(\V{w}^{(1)} - \V{w}^{(2)})}{4}
\left[1+\tanh{\left(\frac{y-9.0\times 10^{-6}}{5.625\times 10^{-7}}\right)}\right]
\left[1+\tanh{\left(\frac{2.7\times 10^{-5}-y}{5.625\times 10^{-7}}\right)}\right].
\end{equation}

\subsubsection{Convergence Test}
We perform a deterministic convergence test using the initial conditions described
above.  We perform simulations using $128^2$, $256^2$, $512^2$, and $1024^2$ grid cells
($\Delta x \approx 2.81, 1.41, 0.70, 0.35$~nm),
with corresponding time steps of $\Delta t =$0.1, 0.05, 0.025, and 0.0125~ns.
We note that for the coarsest simulation we are very close to the electrostatic
stability limit of 0.116~ns, and for the finest simulation we are very close to
the mass diffusion stability limit of 0.0151~ns.
We run each simulation to 10~ns.
We compute the error in each simulation by comparing it to coarsened
data from the next-finer simulation.
In Table \ref{ConvergenceTable}, we show the $L^1$ norm errors
and convergence rates for the total density,
concentrations, charge, and $y$-velocity at the final time.  As expected, the
method is clearly second-order in all variables.
\begin{table}
  \centering
  \begin{tabular}{|c||c|c|c|c|c|c|}
     \hline
     Variable & $L_{64-128}^1$ Error & Rate & $L_{128-256}^1$ Error & Rate & $L_{256-512}^1$ Error \\
     \hline
$\rho$ & 6.01e-15 & 2.01 & 1.49e-15 & 2.01 & 3.71e-16 \\
$w_1$ & 3.38e-15 & 2.01 & 8.37e-16 & 2.00 & 2.09e-16 \\
$w_2$ & 5.22e-15 & 2.02 & 1.29e-15 & 2.00 & 3.22e-16 \\
$w_3$ & 8.59e-15 & 2.01 & 2.13e-15 & 2.00 & 5.31e-16 \\
$q_f$ & 2.90e-07 & 1.97 & 7.38e-08 & 2.00 & 1.85e-08 \\
$v$ & 1.09e-13 & 2.02 & 2.69e-14 & 1.99 & 6.77e-15 \\
     \hline
   \end{tabular}
  \caption{$L^1$ errors comparing successively refined solutions and convergence
           rate for the diffusing saltwater deterministic example.
           Similar convergence rates are obtained for other norms.}
\label{ConvergenceTable}
\end{table}

\subsubsection{Electroneutral Approximation}

In order to see the effect of including charged species as compared to charge-neutral fluids,
we consider the coarsest resolution setup from the previous section.
We now run the $128^2$ simulation using the same $\Delta t = 0.1$~ns, but
to a final time of 1~$\mu$s.  We run a second simulation with the exact same
configuration, but set $\V{z}=0$.  Finally, we run a third simulation with
the exact same configuration, but set $\V{z}=0$ and modify the
self-diffusion coefficients
of both ions to be equal to the effective diffusion coefficient \eqref{ambipolarCoeff}, which here reduces to
\begin{equation}
D_{\rm amb} = \frac{2D_{\rm Na} D_{\rm Cl}}{D_{\rm Na} + D_{\rm Cl}} \approx 1.61\times 10^{-5}~\textrm{cm$^2$/s},
\end{equation}
which is what the electroneutral approximation
gives as the apparent diffusion coefficient of NaCl \cite{grodzinsky2011field}.

\begin{figure}
  \centering
  \includegraphics[width=3.2in]{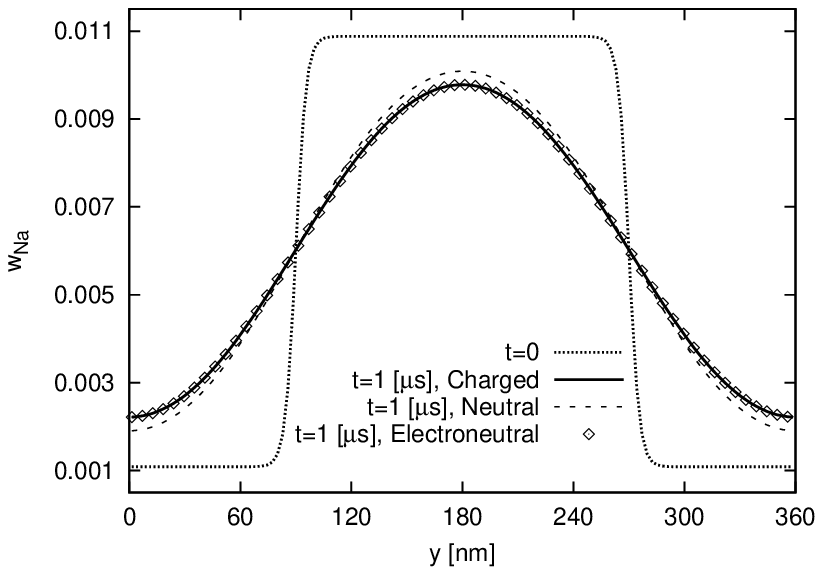} \includegraphics[width=3.2in]{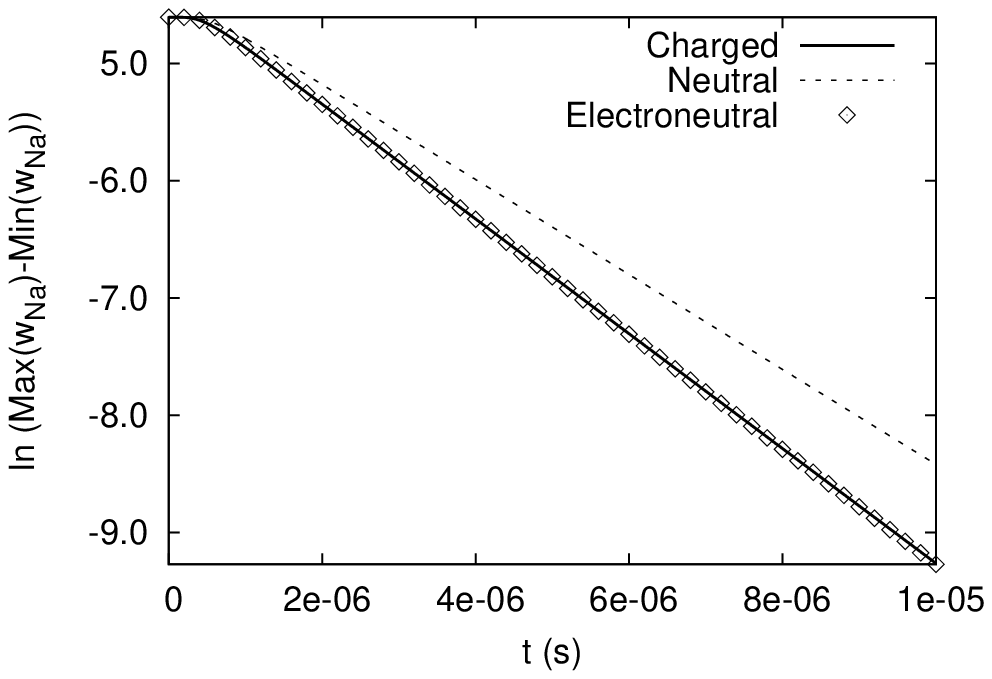}
  \caption{A comparison of diffusing saltwater with and without the effect
           of charges, as well as a comparison to the electroneutral approximation.
           A strip of saltwater is placed in the center of the domain and allowed to diffuse
           deterministically.  (Top) The initial configuration, and later-time
           profiles for the three simulations.  (Bottom) The logarithm of the difference between the maximum and minimum
           of the concentration versus time.  The slopes of these lines are consistent with the effective
           diffusivities used in the neutral and electroneutral cases.}
  \label{DetermProfile}
\end{figure}
In the left frame of Figure \ref{DetermProfile}, we plot the initial configuration
of $w_{\rm Na}$ as a function of $y$, and the final configurations for the three
simulations. \modified{For a problem in which the problem domain is a factor of $\sim$250 larger than the Debye length $\lambda_D^{(2)}$,}
we can graphically see that the electroneutral approximation matches the charged
species code, but not the charge-neutral simulation.
In the right frame of Figure \ref{DetermProfile}, we show the
difference between the peak values in the three simulations as a function of time,
and show how the peak values over time are consistent with the effective
diffusivities.  In particular, for the neutral and electroneutral cases, we know that at late times this
logarithmic quantity decays linearly in time, with a slope proportional to the diffusivity. In the charged
case, this behavior is preserved, with an ``effective'' diffusivity within 0.25\% of that given by the
electroneutral model. The values of the diffusivity extracted from the slopes, 1.33 $\times 10^{-5}$ cm$^2$/s 
in the neutral case and 1.61 $\times 10^{-5}$ cm$^2$/s in the electroneutral case, are in agreement 
with the simulation parameters.

\subsection{Stochastic Tests}

\subsubsection{Equilibrium Structure Factor}
\begin{figure}
  \centering
  \includegraphics[width=3in]{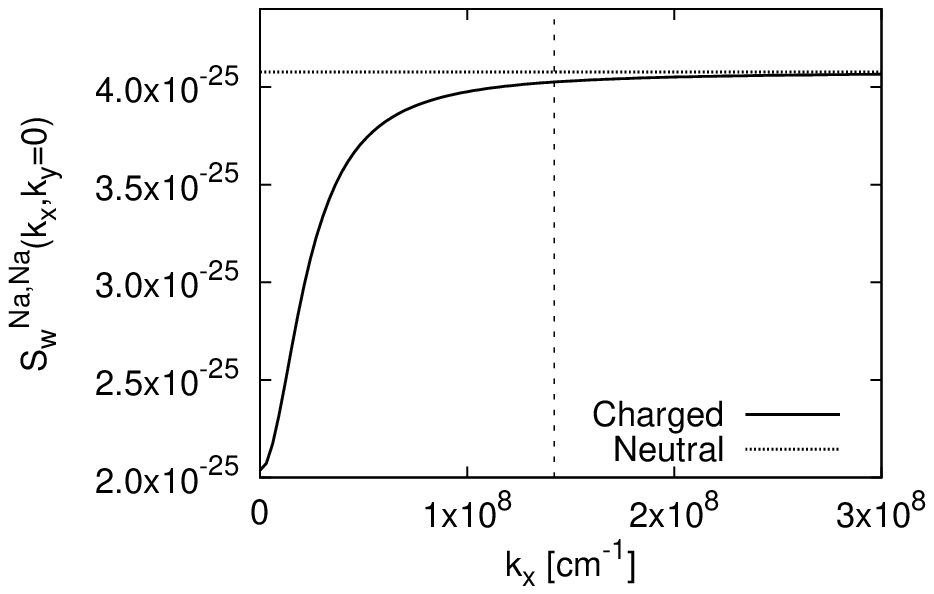}
  \includegraphics[width=3in]{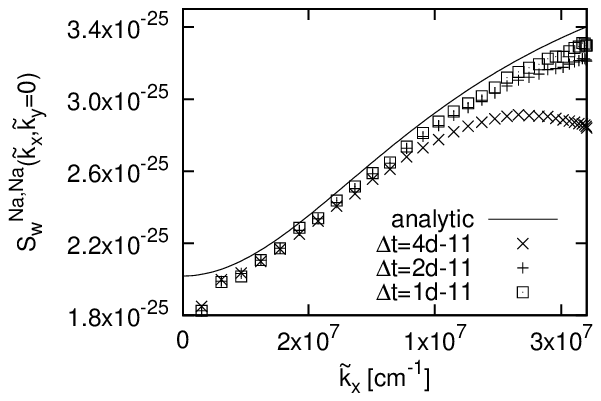}\\
  \includegraphics[width=3in]{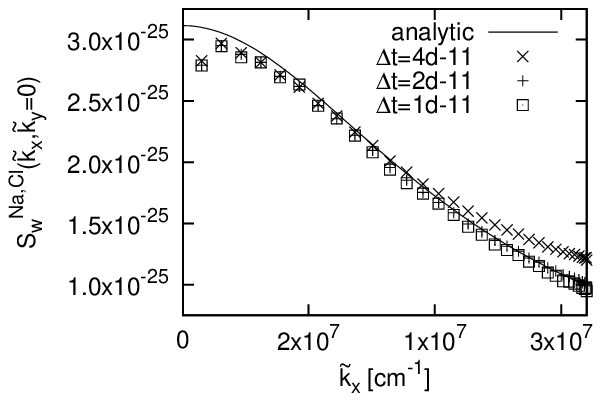}
  \includegraphics[width=3in]{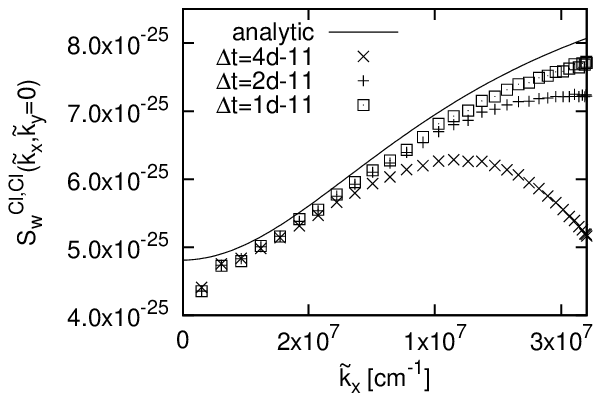}
  \caption{(Top-Left) Analytical static structure factor for Na-Na correlation in saltwater,
            $S_w^{\rm{Na,Na}}(k_x,k_y=0)$.  The vertical bar correspond to the Debye wavenumber.
            The structure factor obtained with uncharged species is shown for reference.
            (Top-Right, Bottom) Computed and analytical discrete static structure factors, $S_w^{ij}(\widetilde{k}_x,\widetilde{k}_y=0)$ 
             in saltwater for Na-Na correlation; Na-Cl correlation; Cl-Cl correlation.}
  \label{Saltwater_StructFig}
\end{figure}

We now perform equilibrium simulations of the structure factor
in two dimensions and compare
to theory.  Our initial state is uniform everywhere
given by $\V{w}^{(1)}$ in (\ref{eq:w1}); other fluid parameters are given in
Table~\ref{ParametersTable}.
The Debye length is $\lambda_D = 4.42\times 10^{-8}$~cm, which corresponds to a
Debye wavenumber of $k_D = 2\pi/\lambda_D \equiv 1.42\times 10^8$ cm$^{-1}$.
In Figure \ref{Saltwater_StructFig}, we plot the analytical structure
factor for the Na-Na correlation, and include the Debye wavenumber as a reference.
The structure factor is relatively constant for wavenumbers larger than $k_D$.

For our simulation, the two-dimensional domain is
square with length $L = 4\times 10^{-6}$~cm \modified{ and periodic boundary conditions.
 For these tests, we adapted the variance of the fluctuations so that the non-linear simulation operates 
in the linear regime assumed for deriving the structure factors in section~\ref{sec:Structure}.}
The simulation box has $64\times 64$ grid cells ($\Delta x = 6.25\times 10^{-8}$~cm)
and the time step sizes are
$\Delta t = 1\times 10^{-11}, 2\times 10^{-11}$, or $4\times 10^{-11}$~s.
The largest time step we used corresponds to $\sim$ 80\% of the explicit mass
diffusion stability limit, and is $\sim$ 400 times the explicit viscous stability limit,
(recall that we treat mass diffusion explicitly and viscosity implicitly).
We skip the first $10^5$ time steps and then
collect samples from the subsequent $9\times 10^5$ steps.

When comparing against continuum theory, we account
for errors in the discrete approximation to the continuum Laplacian by using the
{\em modified wavenumber} \cite{LLNS_Staggered},
\begin{equation}
\widetilde{k}_x = k_x\frac{\sin(k_x\Delta x/2)}{k_x\Delta x/2},\label{eq:ktilde}
\end{equation}
instead of the unmodified wavenumber $k_x$.

Figure \ref{Saltwater_StructFig} shows
the structure factors 
for the Na-Na, Na-Cl, and Cl-Cl correlations as a function of $\tilde{k}_x$ given $\tilde{k}_y = 0$.  These are essentially horizontal
profiles about the centerline of the two-dimensional structure factors.
The structure factors approach the analytic solution as the
time step is reduced.
However, at the same time, we also see significant errors
at the larger wavenumbers as the time step approaches the stability limit,
as expected for any explicit time stepping method \cite{LLNS_S_k}.

\begin{figure}
  \centering
  \includegraphics[width=3.2in]{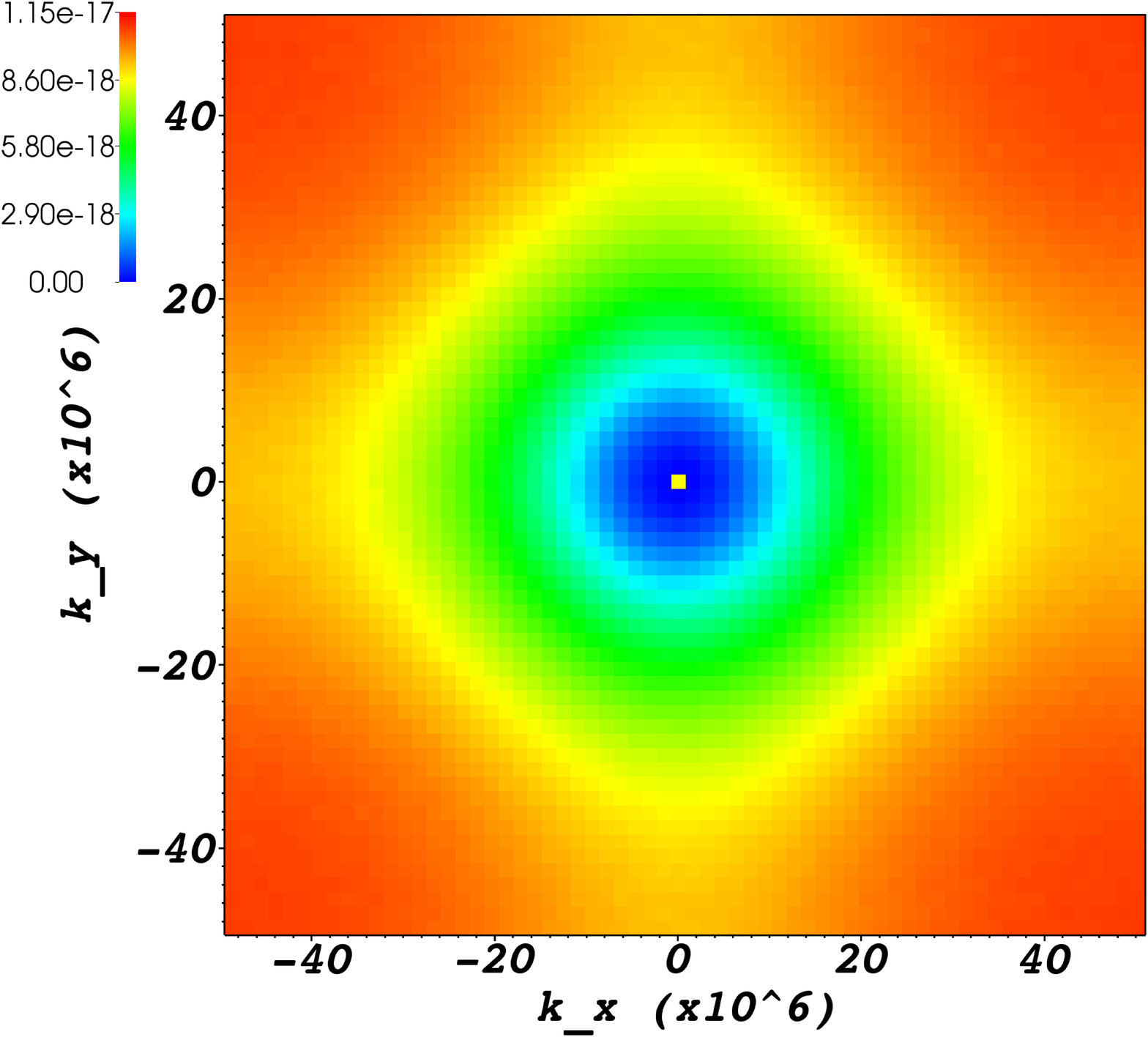}
  \includegraphics[width=3.2in]{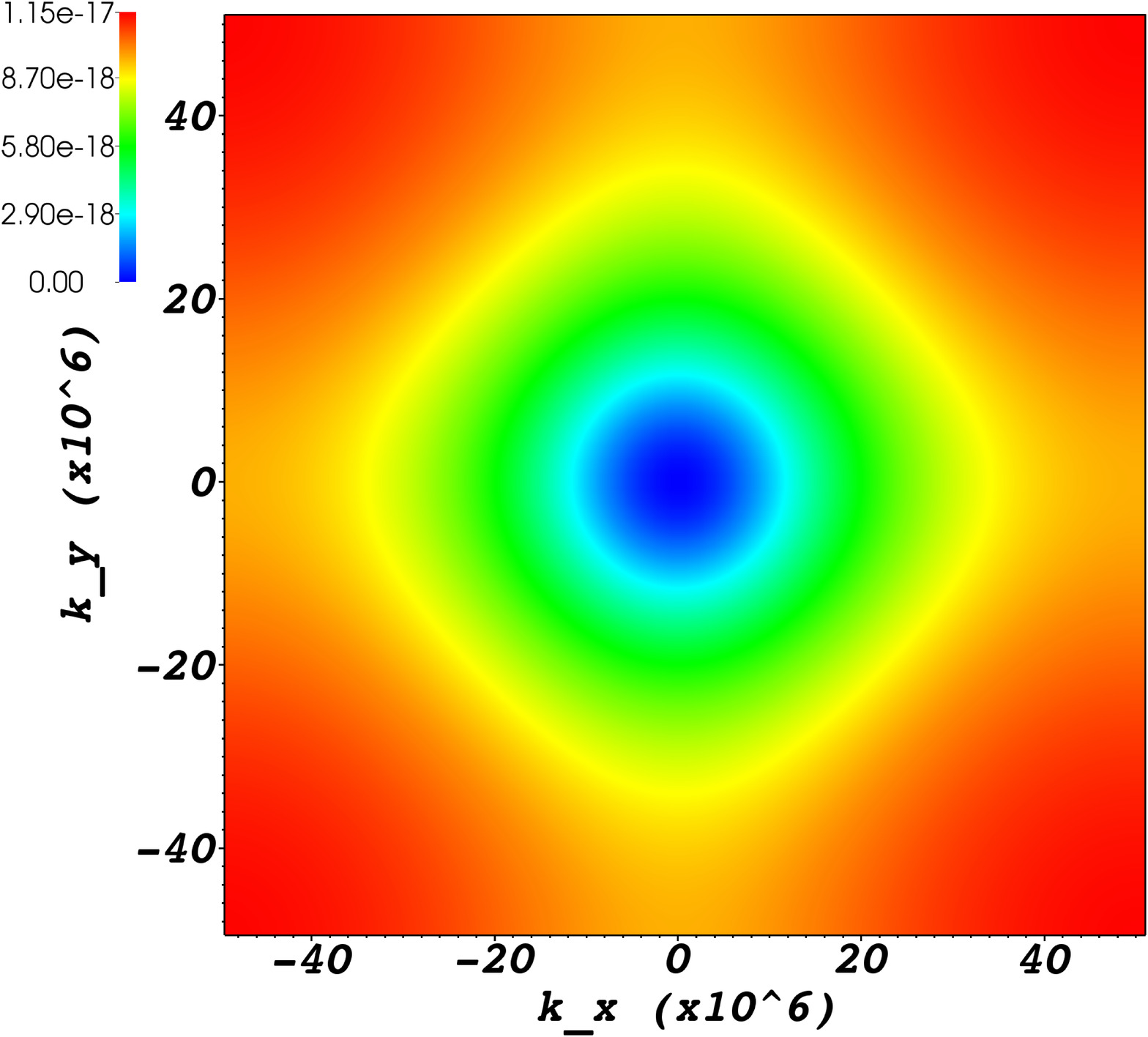}
  \caption{\modified{The (left) computed and (right) analytical discrete static 
    structure factor,
    $S_{\bar{z}}(\V{k})$ in saltwater for the specific charge $\bar{z}$ for
    $\Delta t = 1\times 10^{-11}$~s.
    Note that the lack of perfect rotational isotropy comes from the 
    spatial discretization errors and the fact that the axes here show 
    the unmodified wavenumber $\V{k}$ rather than $\widetilde{\V{k}}$.
    }}
  \label{Saltwater_StructFig2d}
\end{figure}

Figure \ref{Saltwater_StructFig2d} shows the predicted and measured
structure factors as a function of 
$\V{k} = (k_x,k_y)$ 
\modified{for the specific charge $\bar{z}=\sum_{k=1}^{N} w_k z_k$} for saltwater for the smallest time step.  The agreement 
between the analytical and computed structure factors is excellent.

\subsubsection{Nonequilibrium Giant Fluctuations}\label{GiantFluctSection}

In this section we analyze a system that is out of equilibrium, following the approach of 
Refs. \cite{MultispeciesCompressible,LowMachMultispecies}.
We first simulate the evolution in time of saltwater (see Table~\ref{ParametersTable} and previous section) in a two-dimensional square domain of side length $L=3.2\times 10^{-5}$ cm.
We use 64$\times$64 cells ($\Delta x = 5\times 10^{-7}$ cm).
The two side boundaries are periodic, while the top and bottom boundaries are fixed reservoir boundaries for mass fractions, with
respective mass fractions given by (\ref{eq:w1}) and (\ref{eq:w2}).
A relatively small time step of $10^{-10}$~s (less than 5\% the diffusive stability limit) was used in these calculations to ensure that the temporal integration errors are smaller than the statistical errors.

After waiting for a sufficiently large number of time steps for the fluctuations to become statistically stationary,
we calculate the Fourier spectrum $\widehat{\delta w}_i(k_x)$ of the fluctuations of the mass fractions
averaged along the gradient (vertical averages).
We then calculate the associated structure factor
$S_w^{ij}(k_x) = \langle  \widehat{\delta w}_i(k_x), \widehat{\delta w}_j(k_x) \rangle$.
From the seawater parameters that we are using, the Schmidt number is
larger than 500, which allows us to assume that the velocity dynamics is overdamped (see Section \ref{StructFactGiantFluct}).

\begin{figure}
  \centering
  \includegraphics[width=3.2in]{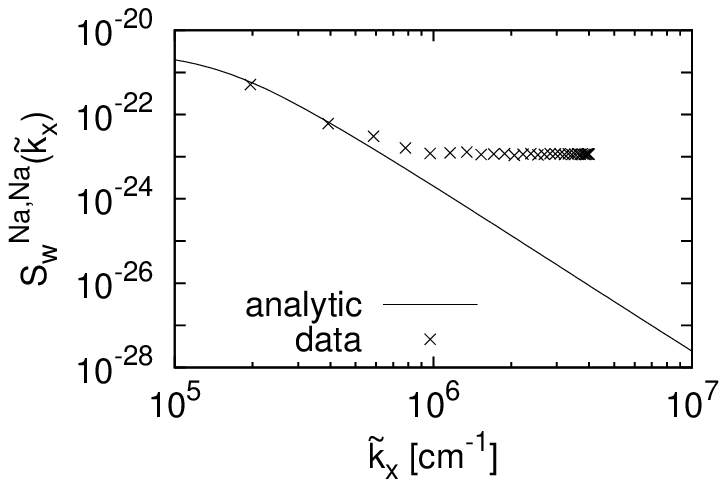}
  \includegraphics[width=3.2in]{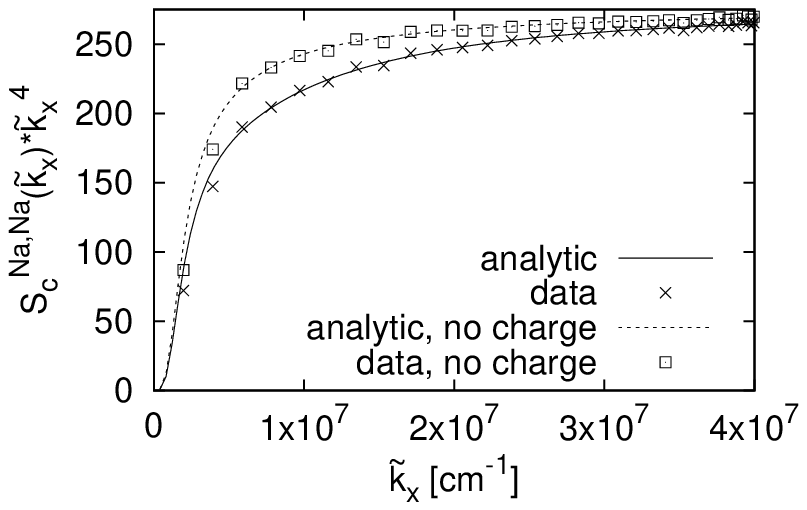}\\
  \includegraphics[width=3.2in]{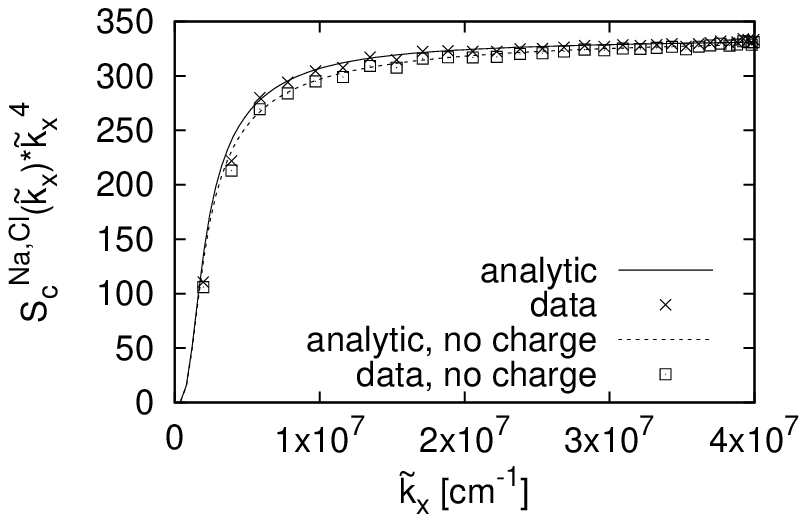}
  \includegraphics[width=3.2in]{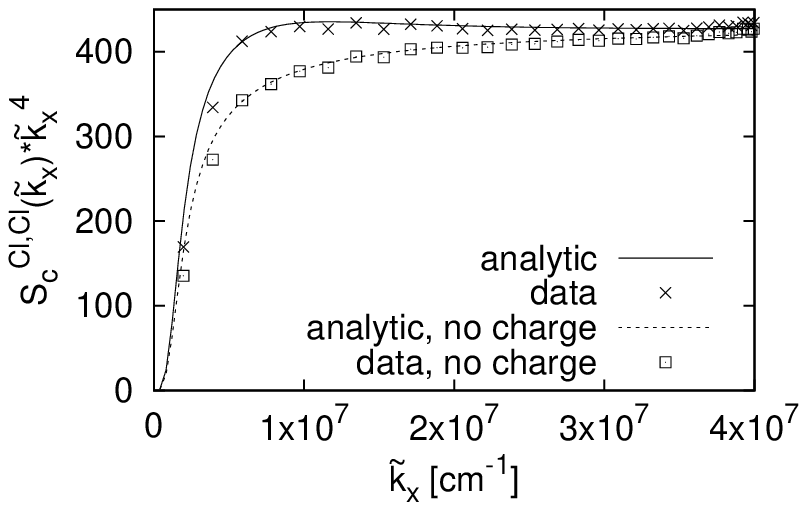}
  \caption{(Top-Left) Amplitude of the fluctuations of the vertically averaged mass
           fraction of Na.
           (Top-Right and Bottom) Structure factors $S_c$ of the vertical averages,
           multiplied by $\widetilde{k}_x^4$, where $\widetilde{k}_x$ is given by
           (\ref{eq:ktilde}). For comparison, the case where charges effect are ignored
           is also represented.}
  \label{GiantFluctStruct}
\end{figure}

In the top-left of Figure \ref{GiantFluctStruct},
we show the amplitude of the fluctuations.
Although a slope approaching -4 in logarithmic
scale is visible for small wavenumbers, the amplitude plateaus for large wavenumbers.
This is due to the equilibrium fluctuations in the mass fractions, which, due to the relative weakness of
the mass fraction gradients, tend to hide the nonequilibrium contribution to the fluctuations.
We can alleviate this effect by simulating larger systems,
but this is difficult because of our desire to resolve the Debye length.
Instead, we choose to artificially remove the fluctuations in the species
fluxes and only include fluctuations in the momentum flux,
thus giving us the non-equilibrium contribution to the structure factor (see Eq.~(\ref{OUeqnGiantFluct_neq})).
For these simulations we use a domain of side  3.2$\times 10^{-6}$ cm and a time step size of 5 ps.

We note that due to the confinement effect induced by the two reservoirs, the structure factor for small $k_x$ is reduced.
This can be {\em approximately}
accounted for by multiplying the bulk theoretical results
by a confinement factor \cite{deZarate2015GiantFluctFiniteEffects} and defining
\begin{equation} \label{S_k_confined}
\M{S}_{c}(k_x) = \left( 1+\frac{4(1-\cosh(k_x L_y))}{k_x L_y(k_x L_y+\sinh(k_x L_y))} \right) \M{S}_{\textrm{neq}}(k_x),
\end{equation}
where $\M{S}_{c}$ is the corrected quantity accounting for the confinement effect, and $L_y$ is the distance between the two reservoir walls.
Note that this approximation becomes exact in the electroneutral limit, since the equations reduce to those used in \cite{deZarate2015GiantFluctFiniteEffects} for a binary mixture, but with an effective diffusion coefficient for the solute.
Since it is exactly for small wavenumbers (large scales) that the confinement effects are large, we expect the approximation (\ref{S_k_confined})
to be a reasonably good approximation over all wavenumbers; this is confirmed by a comparison to our numerical solution.

Figure \ref{GiantFluctStruct} shows the resulting structure factors $S_{c}^{\rm Na,Na}$, $S_{c}^{\rm{Cl,Cl}}$ and $S_{c}^{\rm{Na,Cl}}$. In order to enable a more accurate visualization of the difference between the theory and the code output, these quantities are multiplied by $\widetilde{k}_x^4$.
To assess the effect of charges, we also plotted the results from the same system where the charges are artificially set to zero.
The difference between the two cases reaches 25\% for the chloride-chloride correlation
with excellent agreement between the numerical results and the theory.

\subsection{Electrostatically Induced Mixing Instability}\label{MixingSection}

In this section, we study the effect of fluctuations and imposed boundary potential
on the
three-dimensional mixing of two layers of water with different initial salinity levels.
The domain is cubic with sides of length $L=4\times 10^{-4}$~cm.
The saltier water is initially on the lower-half of the domain, with lower and
upper concentrations given by (\ref{eq:w1}) and (\ref{eq:w2}).
The initial interface is smoothed slightly in the vertical direction
with a hyperbolic tangent profile over a few grid cells.
We impose periodic boundary conditions on the lateral boundaries, and no-slip
walls on the vertical boundaries with imposed values of electric potential,
$\ElectricPotential$.  The simulation uses $128^3$ grid cells
($\Delta x = 3.125\times 10^{-6}$~cm) with a time step of $5\times 10^{-11}$~s,
which is roughly 50\% of the electrostatic stability limit.

\begin{figure}
  \centering
  \includegraphics[width=3.2in]{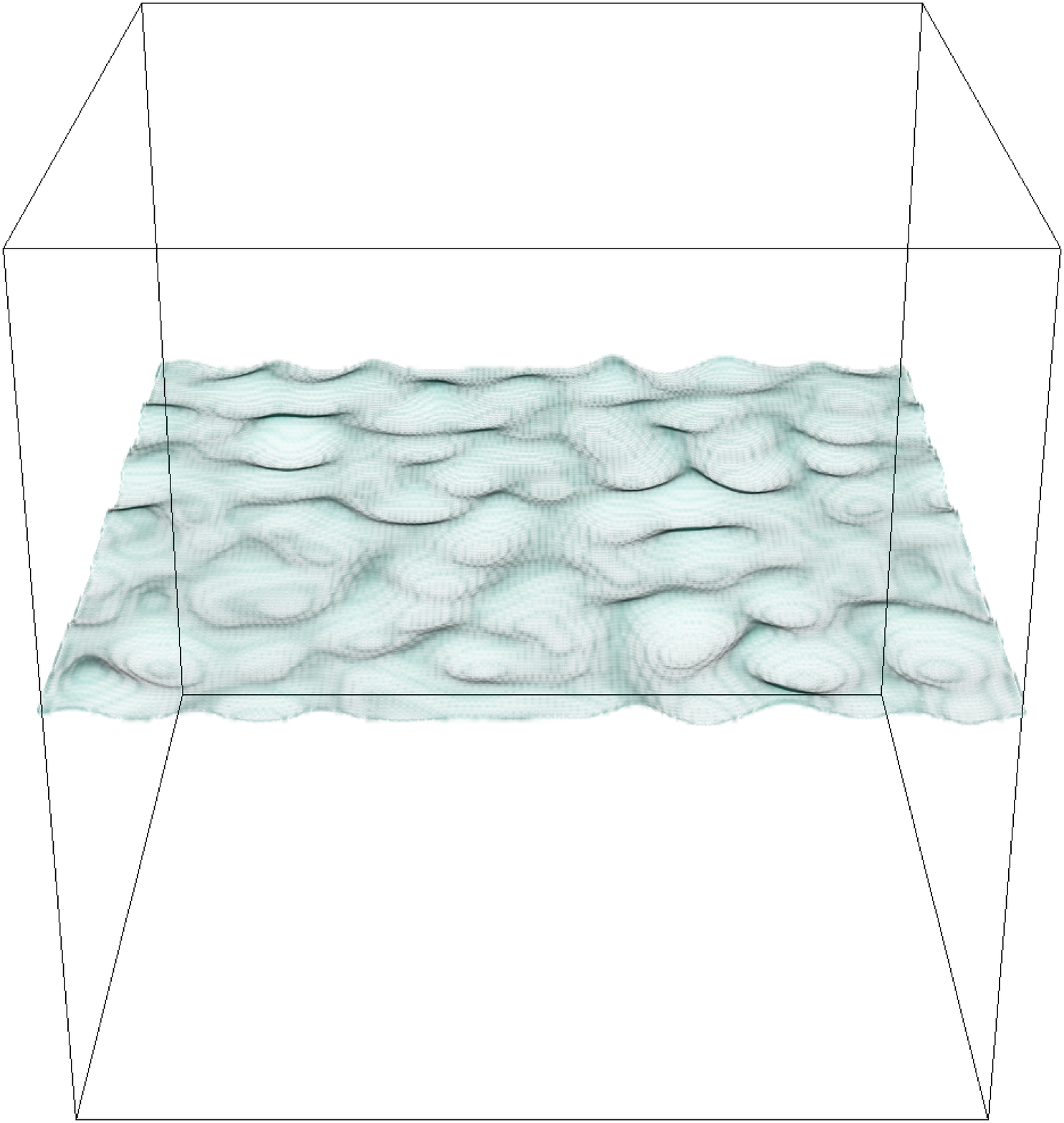}
  \includegraphics[width=3.2in]{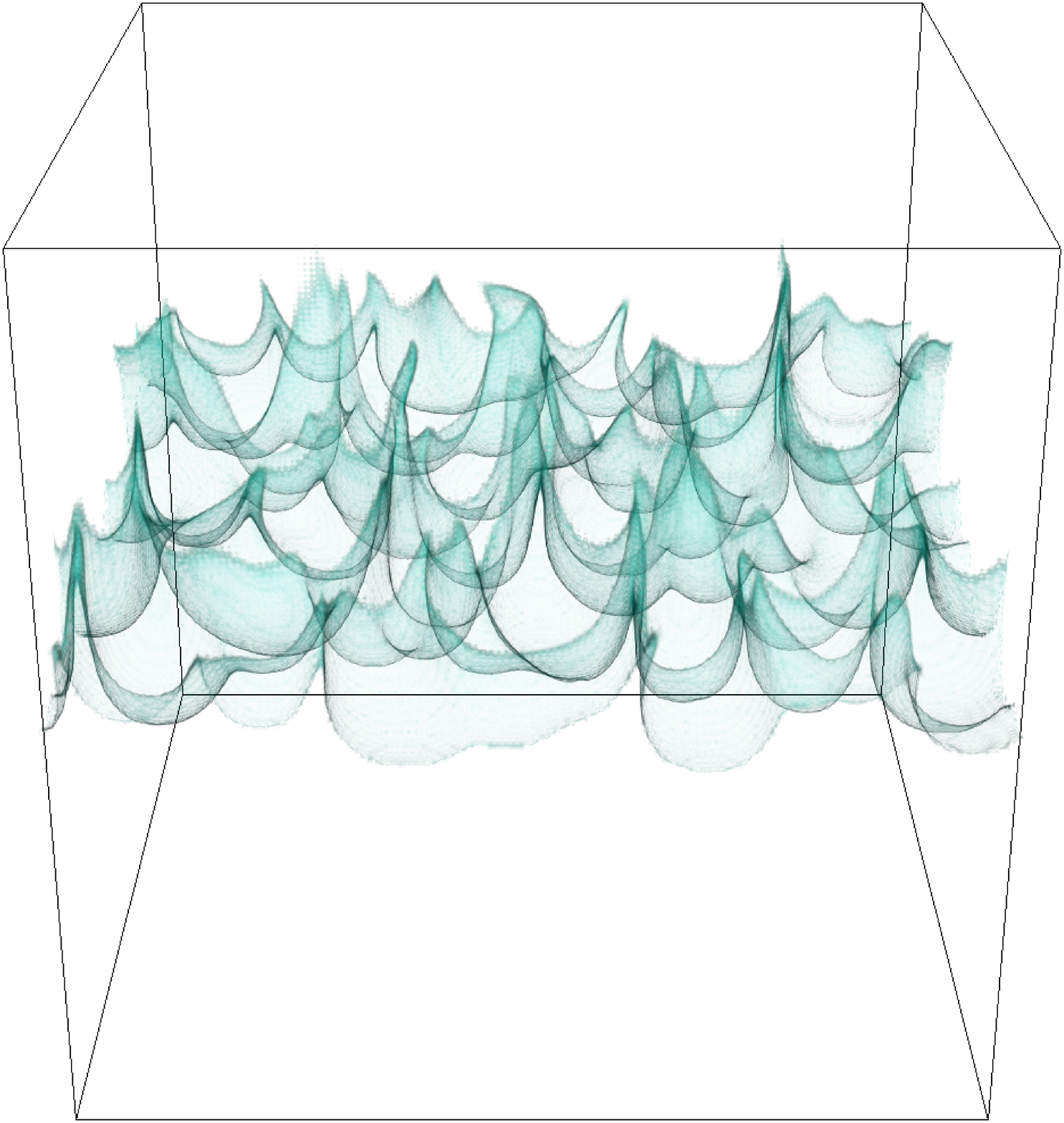}\\
  \includegraphics[width=3.2in]{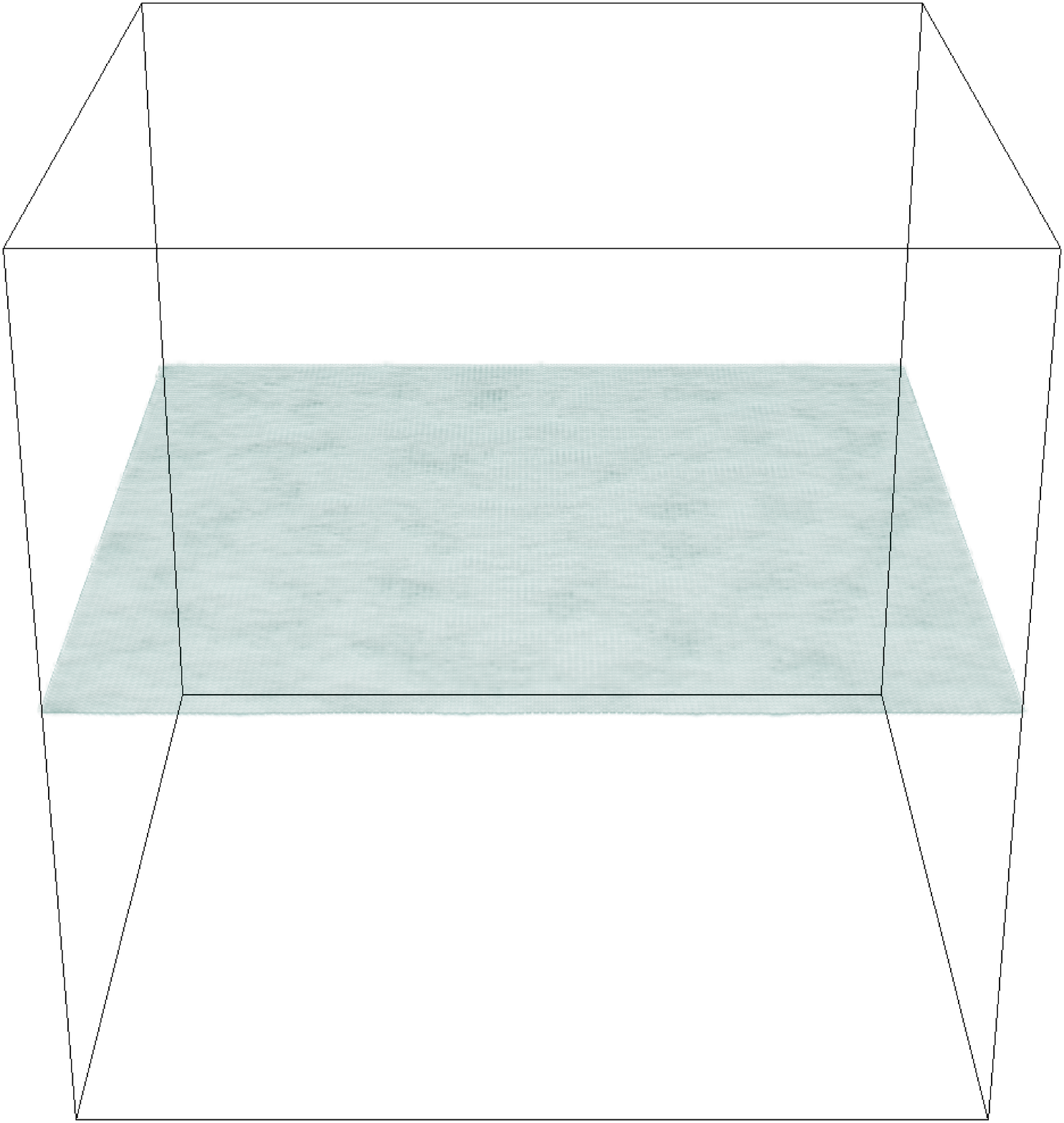}
  \includegraphics[width=3.2in]{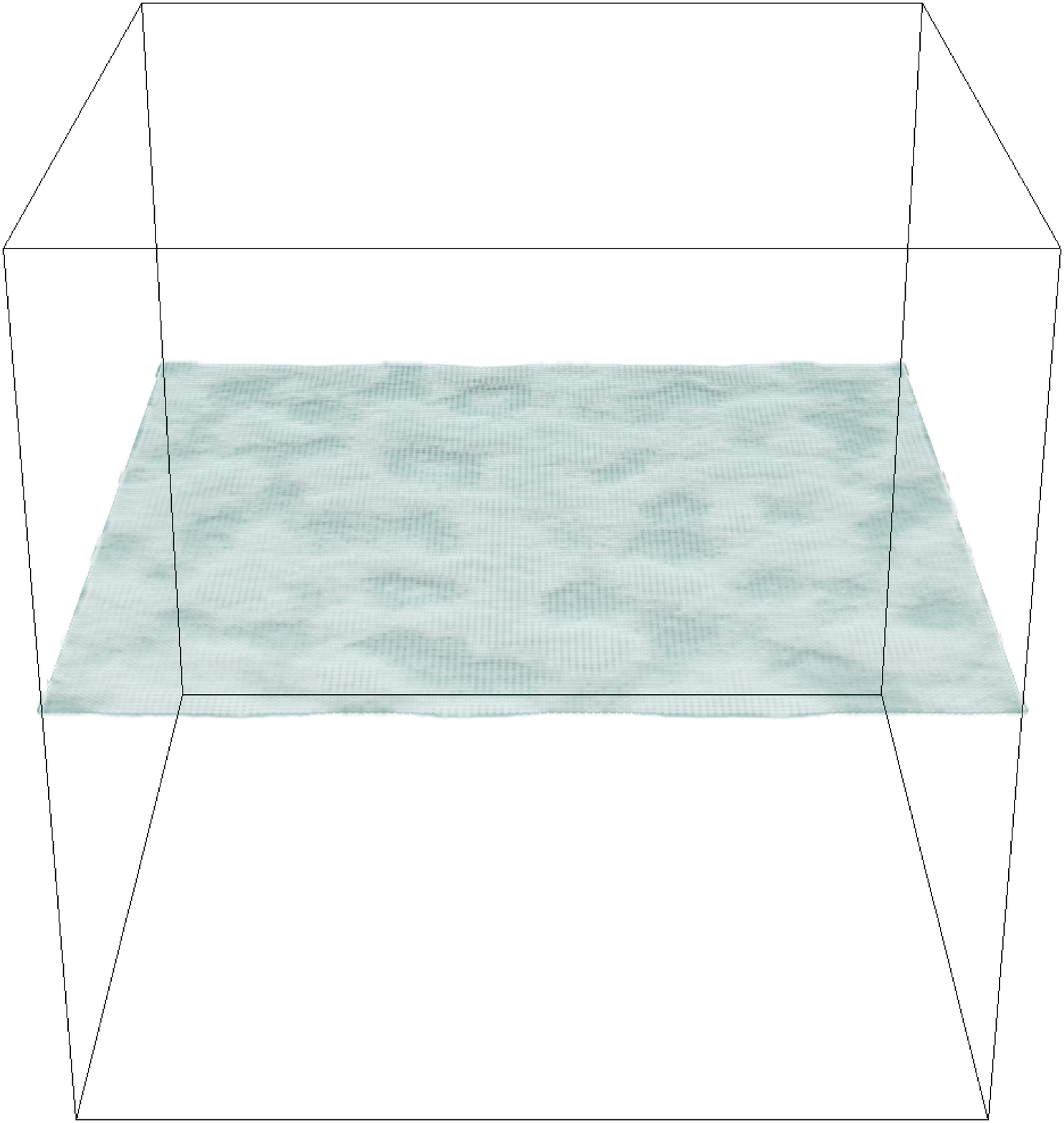}
  \caption{Mixing instability with varying applied potential difference across
           the upper/lower boundaries.  The initial interface is flat.
           Shown is a contour of density with value
           halfway between the density of the saltier water below and fresher
           water above.  The top images
           correspond to $200$~V at $t=5\times 10^{-8}$ and $1\times 10^{-7}$~s.
           The bottom images
           correspond to $100$~V at $t=5\times 10^{-8}$ and $1\times 10^{-7}$~s.
           }
  \label{Potential_Instability}
\end{figure}

Here we demonstrate that there is an instability brought on by an imposed potential.
Initially a charge separation forms at the interface due to the difference in the
diffusivities between the two types of ions.
This charge separation happens even without any imposed potential.
The interface begins to diffuse with slight roughness caused by fluctuations.
We observe that in simulations with \modified{a sufficiently} large applied potential, the imposed electric
field is strong enough to accelerate charges in localized regions of the interface
toward the vertical walls faster than mass diffusion can smooth the interface,
and an instability develops.
In Figure \ref{Potential_Instability}, we show snapshots
at two different times from simulations with $200$~V and $100$~V
potential difference across the boundaries.
In the $100$~V case, the instability does not develop, whereas in the $200$~V case
we see significant interface deformation.
In a forthcoming paper we will provide detailed analysis on the growth of various
spectral modes
in the presence of different strength electric fields, with/without the inclusion
of thermal fluctuations.

\section{Conclusions and Future Work}

We have developed a low Mach number fluctuating hydrodynamics formulation for mixtures
of charged species suitable for modeling electrolyte mixtures.
The model and algorithms are based on
the ones previously developed in
\cite{LowMachExplicit, LowMachImplicit, LowMachMultispecies} combined with
a quasielectrostatic approximation for the effect of localized charges.  We have verified
second-order accuracy in the deterministic setting, and have shown that our code gives
results consistent with the electroneutral approximation for simple diffusion problems
at length scales larger than the Debye length.
We have also verified that our model
can accurately capture static equilibrium \modified{fluctuations, as well as} nonequilibrium
fluctuations in the presence of an imposed concentration gradient.  Our model predicts
an instability between layers of saltwater and freshwater in the presence
of an applied potential difference. This instability is reminiscent of the electrokinetic
instabilities studied in \cite{Lin2004,Posner2012}. The detailed dynamics of the instability
and the impacts of fluctuations on it and on the effective diffusion coefficient will be the
subject of future investigation.

\modified{As mentioned in Section~\ref{ModelSection}, our formulation assumes that one can separate the contributions of the long-ranged Coulomb electrostatic interactions from those due to short-ranged molecular interactions. This is in many ways similar to the assumptions needed to justify fluctuating Ginzburg-Landau models for multiphase liquid mixtures \cite{CHN_Compressible,FluctDiff_FEM}, namely, that surface tension arises out of a long-ranged attractive potential acting on scales larger than the short-ranged repulsion (see Appendix A in \cite{FluctDiff_FEM}). We therefore implicitly assume that the Debye length is substantially larger than the molecular scale. More specifically, to justify the equations written here from the theory of coarse graining one would assume that each coarse graining volume \cite{FluctDiff_FEM} (hydrodynamic cell in our discretization) contains many molecules, and can be described using thermodynamic potentials (notably, chemical potentials) that refer to the fluid mixture in the {\em absence} of electrostatics. This allows us to consider solvents that are themselves non-ideal mixtures, and to capture some molecular effects such as solvation layers
in the thermodynamic potentials {\em before} electrostatics is accounted for. It is important to note that, unlike the majority of theoretical work on electrolytes \cite{Electrolytes_DH_review}, we do {\em not} assume that the solvent-solute mixture would be ideal if there were no electrostatics.}

In Section \ref{StructureFactor} we derived a completely general theory for the
equilibrium fluctuations in an electrolyte at equilibrium and in the presence of a concentration gradient.
Unlike existing results in the literature we did not assume dilute or ideal solutions, and
considered an arbitrary number of solvent and solute species. To our knowledge, existing
literature does not properly define the Debye length for a nonideal mixture.
Our results describe the fluctuations of concentrations at scales below the Debye length. As such,
they are unlikely to be accessible to experimental confirmation via light scattering or other
techniques traditionally used to study fluctuations. Nevertheless, the analytical results are important in interpreting
results from molecular dynamics simulations \cite{MS_diffusion_ionic}
aimed at measuring the transport coefficients from mesoscopic fluctuations \cite{TransportThermo_MD}.

\modified{In Section \ref{ExamplesSection} we studied a number of examples involving seawater, 
which is a not-so-dilute solution with a rather small value of the Debye length, $\lambda_D\sim1$~nm. 
The lack of clear separation between the Debye length and the molecular scale might put into question 
the validity of the fluctuating hydrodynamics (FHD) approach we have used in this work. 
However, there are a number of reasons to be optimistic about the usefulness of FHD. 
Firstly, the average number of electrolyte molecules in a Debye volume is $N_D \sim x  n\lambda_D^3 \sim x^{-1/2}$, 
where $x$ is the mole fraction of either electrolyte. Therefore, for more dilute solutions $N_D\gg1$ and the 
fluctuating continuum level of description is more appropriate. Secondly, we note that there is supporting 
evidence that fluctuating hydrodynamics is useful as a {\em discrete} coarse-grained description at very small scales, 
well beyond what can be justified mathematically. For example, comparisons to molecular dynamics (MD) 
simulations \cite{StagerredFluct_Inhomogeneous} have shown that FHD provides a surprisingly accurate 
description of fluid interfaces, even though the thickness of the interface is only a couple of 
nanometers and each hydrodynamic cell contains less than ten water molecules. 
In Section~\ref{ModelSection}, we demonstrated that our analytical results reproduce the well-known 
Debye-Huckel theory for dilute solutions. As detailed in the review article~\cite{Electrolytes_DH_review}, 
DH theory is known to be surprisingly successful in describing solutions well beyond the ideal dilute 
regimes in which it can be justified, however, the coefficients appearing in the equations must be taken 
as {\em effective} charges and screening lengths. We expect that a similar conclusion applies to FHD: 
If the various transport and thermodynamic quantities are suitably {\em renormalized} based on the hydrodynamic 
cell size (coarse-graining length), perhaps using a direct comparison to MD (see Appendix C 
in \cite{LowMachExplicit} for an illustration), hydrodynamics can efficiently account for the long-ranged 
and long-lived effects that cannot be captured in a reasonable computational effort in direct molecular simulations.}

\deleted{In this work, we showed examples in which using the electroneutral approximation at large length scales
could be a way to sidestep the timescale constraints that result from the full dynamics.}
\modified{We showed that at large length scales, the deterministic part of our algorithm is consistent with results
obtained when the electroneutral approximation is used. By replacing the electric potential term in the diffusion
equations with effective diffusion terms, the electroneutral approximation lifts the timestep stability constraint \eqref{dt_electro_dilute}
induced by the electric term. For solutions such as seawater, the electroneutral approximation holds at length scales where
fluctuations are not negligible (100 nm - 1$\mu$m). Yet, a theory on how to treat the fluctuations of the charged 
species within this approximation has yet to be developed.}
At experimental scales much larger than the Debye length, the fluctuations described by our theory
renormalize the thermodynamic and transport properties entering in the electroneutral/ambipolar approximation.
Imposing the electroneutral
approximation in the context of fluctuating hydrodynamics requires projecting the stochastic fluxes on the
electroneutral constraint, as we will study in future work.


We are developing an implicit discretization for the electric potential driving force
in the mass equations that will allow for longer time integration for problems with
larger length scales, or for mixtures with smaller Debye length.
Additionally, we will expand the extent of physical phenomena
that are accounted for in our method. First of all, while our previous work for neutral multispecies
mixtures included the use of the energy equation in order to deal in particular with temperature gradients,
we chose here to consider only isothermal systems because we wanted to limit the number of physical
phenomena and parameters that might affect the systems of interest. Including the energy equation will be
a direct extension and should not present conceptual difficulties at this stage. At a similar level, the
equation of state~\eqref{EOS_Eqn} that we use can be generalized to more realistic models.

\modified{The isothermal low Mach number model neglects the contributions of barodiffusion
and thermodiffusion.
While this is a good approximation for most practical problems, omitting barodiffusion is not strictly consistent with equilibrium statistical mechanics. This is because barodiffusion has thermodynamic rather than kinetic origin and is responsible for effects such as gravitational sedimentation (see Appendix B in \cite{LowMachMultispecies}). Corrections to sedimentation profiles due to electrostatic contributions of the osmotic pressure may not be correctly captured in the present formulation; we will explore these issues in future work.}

The permittivity is assumed to be constant in our simulations.
In reality, however, the relative permittivity of a mixture depends
on concentration. For example, for sea water it is about 7 percent lower
than in fresh water, so neglecting these variations is not
entirely justified for the simulations presented in Section~\ref{MixingSection}.
Furthermore, the dielectric nature of water results in physical phenomena such
as polarization charges and polarization currents \cite{grodzinsky2011field} whose behaviors are unclear
from the standpoint of fluctuating hydrodynamics and which represents an exciting
direction for future research.

In the longer term we would like to incorporate more realistic microscopic models for electrolyte behavior.
Electrolyte transport is known to be affected by a range of non-linear phenomena,
such as the electrophoretic effect or the Debye-Onsager relaxation effect~\cite{robinson2012electrolyte}.
Simulation of electrochemical processes requires the incorporation of chemical reactions into the models.
Molecular-scale boundary-specific effects play an important role in many cases, as in the simulation of membranes.
Including these types of phenomena may require hybrid algorithms that couple different types of physical models
and algorithms, such as coupling a molecular simulation to a fluctuating hydrodynamics solver (e.g., see~\cite{DSMC_Hybrid}).

\newpage

\bibliographystyle{plain}
\bibliography{References}

\begin{thebibliography}{10}

\bibitem{ABRZ:I}
A.~S. Almgren, J.~B. Bell, C.~A. Rendleman, and M.~Zingale.
\newblock {Low Mach Number Modeling of Type Ia Supernovae. I. Hydrodynamics}.
\newblock {\em \apj}, 637:922--936, February 2006.
\newblock paper I.

\bibitem{MultispeciesCompressible}
K.~{Balakrishnan}, A.~L. {Garcia}, A.~{Donev}, and J.~B. {Bell}.
\newblock Fluctuating hydrodynamics of multispecies nonreactive mixtures.
\newblock {\em Phys. Rev. E}, 89:013017, 2014.

\bibitem{berne_pecora}
B.~J. Berne and R.~Pecora.
\newblock {\em Dynamic Light Scattering}.
\newblock Robert E. Krieger Publishing Company, 1990.

\bibitem{MultispeciesChemistry}
A.~K. {Bhattacharjee}, K.~{Balakrishnan}, A.~L. {Garcia}, J.~B. {Bell}, and
  A.~{Donev}.
\newblock Fluctuating hydrodynamics of multispecies reactive mixtures.
\newblock {\em J. Chem. Phys.}, 142(22):224107, 2015.

\bibitem{bograchev2013nanowires}
D.~A. Bograchev, V.~M. Volgin, and A.~D. Davydov.
\newblock Simulation of inhomogeneous pores filling in template
  electrodeposition of ordered metal nanowire arrays.
\newblock {\em Electrochimica Acta}, 112:279--286, 2013.

\bibitem{StokesKrylov}
M.~Cai, A.~J. Nonaka, J.~B. Bell, B.~E. Griffith, and A.~Donev.
\newblock {Efficient Variable-Coefficient Finite-Volume Stokes Solvers}.
\newblock {\em Comm. in Comp. Phys. (CiCP)}, 16(5):1263--1297, 2014.

\bibitem{MixedDiffusiveInstability}
J.~Carballido-Landeira, P.~M.~J. Trevelyan, C.~Almarcha, and A.~De~Wit.
\newblock Mixed-mode instability of a miscible interface due to coupling
  between rayleigh-taylor and double-diffusive convective modes.
\newblock {\em Physics of Fluids}, 25(2):024107, 2013.

\bibitem{DivergentMaxStefanDiff}
B.~Chakraborty, J.~Wang, and J.~Eapen.
\newblock {Multicomponent diffusion in molten LiCl-KCl: Dynamical correlations
  and divergent Maxwell-Stefan diffusivities}.
\newblock {\em Phys. Rev. E}, 87:052312, May 2013.

\bibitem{CHN_Compressible}
A.~Chaudhri, J.~B. Bell, A.~L. Garcia, and A.~Donev.
\newblock Modeling multiphase flow using fluctuating hydrodynamics.
\newblock {\em Phys. Rev. E}, 90:033014, 2014.

\bibitem{cussler2009diffusion}
E.~L. Cussler.
\newblock {\em Diffusion: Mass Transfer in Fluid Systems}.
\newblock Cambridge Series in Chemical Engineering. Cambridge University Press,
  2009.

\bibitem{LaminarFlowChemistry}
M.~S. Day and J.~B. Bell.
\newblock {Numerical simulation of laminar reacting flows with complex
  chemistry}.
\newblock {\em Combustion Theory and Modelling}, 4(4):535--556, 2000.

\bibitem{FluctDiff_FEM}
J.A. de~la Torre, P.~Espa{\~n}ol, and A.~Donev.
\newblock {Finite element discretization of non-linear diffusion equations with
  thermal fluctuations}.
\newblock {\em J. Chem. Phys.}, 142(9):094115, 2015.

\bibitem{deZarate2015GiantFluctFiniteEffects}
J.~M.~O. De~Z{\'a}rate, T.~R. Kirkpatrick, and J.~V. Sengers.
\newblock Non-equilibrium concentration fluctuations in binary liquids with
  realistic boundary conditions.
\newblock {\em The European Physical Journal E}, 38(9):1--9, 2015.

\bibitem{DM_63}
S.~R. DeGroot and P.~Mazur.
\newblock {\em Non-Equilibrium Thermodynamics}.
\newblock North-Holland Publishing Company, Amsterdam, 1963.

\bibitem{dill2011molecular}
K~.A. Dill and S.~Bromberg.
\newblock {\em Molecular Driving Forces: Statistical Thermodynamics in Biology,
  Chemistry, Physics, and Nanoscience}.
\newblock Garland Science, 2011.

\bibitem{DSMC_Hybrid}
A.~Donev, J.~B. Bell, A.~L. Garcia, and B.~J. Alder.
\newblock {A hybrid particle-continuum method for hydrodynamics of complex
  fluids}.
\newblock {\em SIAM J. Multiscale Modeling and Simulation}, 8(3):871--911,
  2010.

\bibitem{DiffusionJSTAT}
A.~Donev, T.~G. Fai, and E.~Vanden-Eijnden.
\newblock {A reversible mesoscopic model of diffusion in liquids: from giant
  fluctuations to Fick's law}.
\newblock {\em Journal of Statistical Mechanics: Theory and Experiment},
  2014(4):P04004, 2014.

\bibitem{LowMachMultispecies}
A.~Donev, A.~J. Nonaka, A.~K. Bhattacharjee, A.~L. Garcia, and J.~B. Bell.
\newblock {Low Mach Number Fluctuating Hydrodynamics of Multispecies Liquid
  Mixtures}.
\newblock {\em Physics of Fluids}, 27:037103, 2015.

\bibitem{LowMachExplicit}
A.~Donev, A.~J. Nonaka, Y.~Sun, T.~G. Fai, A.~L. Garcia, and J.~B. Bell.
\newblock {Low Mach Number Fluctuating Hydrodynamics of Diffusively Mixing
  Fluids}.
\newblock {\em Communications in Applied Mathematics and Computational
  Science}, 9(1):47--105, 2014.

\bibitem{LLNS_S_k}
A.~Donev, E.~Vanden-Eijnden, A.~L. Garcia, and J.~B. Bell.
\newblock {On the Accuracy of Explicit Finite-Volume Schemes for Fluctuating
  Hydrodynamics}.
\newblock {\em Communications in Applied Mathematics and Computational
  Science}, 5(2):149--197, 2010.

\bibitem{NernstPlanckModel}
W.~Dreyer, C.~Guhlke, and R.~Muller.
\newblock {Overcoming the shortcomings of the Nernst-Planck model}.
\newblock {\em Phys. Chem. Chem. Phys.}, 15:7075--7086, 2013.

\bibitem{LB_StatMech}
B.~D{\"u}nweg, U.~D. Schiller, and A.~J.~C. Ladd.
\newblock {Statistical mechanics of the fluctuating lattice Boltzmann
  equation}.
\newblock {\em Phys. Rev. E}, 76(3):036704, 2007.

\bibitem{GardinerBook}
C.~W. Gardiner.
\newblock {\em {Handbook of stochastic methods: for physics, chemistry \& the
  natural sciences}}, volume Vol. 13 of {\em Series in synergetics}.
\newblock Springer, third edition, 2003.

\bibitem{Electrophysiology_Peskin_Review}
Boyce~E Griffith and Charles~S Peskin.
\newblock Electrophysiology.
\newblock {\em Communications on Pure and Applied Mathematics},
  66(12):1837--1913, 2013.

\bibitem{grodzinsky2011field}
A.~Grodzinsky.
\newblock {\em Field, Forces and Flows in Biological Systems}.
\newblock Taylor \& Francis Group, 2011.

\bibitem{TransportThermo_MD}
S.~Kjelstrup, S.~K. Schnell, T.~J.~H. Vlugt, J.-M. Simon, A.~Bardow,
  D.~Bedeaux, and T.~Trinh.
\newblock Bridging scales with thermodynamics: from nano to macro.
\newblock {\em Advances in Natural Sciences: Nanoscience and Nanotechnology},
  5(2):023002, 2014.

\bibitem{IncompressibleLimit_Majda}
S.~Klainerman and A.~Majda.
\newblock Compressible and incompressible fluids.
\newblock {\em Communications on Pure and Applied Mathematics}, 35(5):629--651,
  1982.

\bibitem{kontturi2008ionic}
K.~Kontturi, L.~Murtom{\"a}ki, and J.~A. Manzanares.
\newblock {\em Ionic Transport Processes: in Electrochemistry and Membrane
  Science}.
\newblock OUP Oxford, 2008.

\bibitem{NegativeMaxStefanDiff1}
G.~Kraaijeveld and J.~A. Wesselingh.
\newblock {Negative Maxwell-Stefan diffusion coefficients}.
\newblock {\em Industrial \& Engineering Chemistry Research}, 32(4):738--742,
  1993.

\bibitem{NegativeMaxStefanDiff2}
G.~Kraaijeveld, J.~A. Wesselingh, and G.~D.~C. Kuiken.
\newblock {Comments on "Negative Maxwell-Stefan Diffusion Coefficients"}.
\newblock {\em Industrial \& Engineering Chemistry Research}, 33(3):750--751,
  1994.

\bibitem{MaxStefElectrolytes}
R.~Krishna.
\newblock Diffusion in multicomponent electrolyte systems.
\newblock {\em The Chemical Engineering Journal}, 35(1):19 -- 24, 1987.

\bibitem{Landau1984electrodynamics}
L.~D. Landau, J.~S. Bell, M.~J. Kearsley, L.~P. Pitaevskii, E.~M. Lifshitz, and
  J.~B. Sykes.
\newblock {\em Electrodynamics of Continuous Media}.
\newblock COURSE OF THEORETICAL PHYSICS. Elsevier Science, 1984.

\bibitem{Landau:Fluid}
L.D. Landau and E.M. Lifshitz.
\newblock {\em Fluid Mechanics}, volume~6 of {\em Course of Theoretical
  Physics}.
\newblock Pergamon Press, Oxford, England, 1959.

\bibitem{LiGregory74}
Y.~Li and S.~Gregory.
\newblock Diffusion of ions in sea water and in deep-sea sediments.
\newblock {\em Geochimica et Cosmochimica Acta}, 38:703--714, 1974.

\bibitem{Lin2004}
H.~Lin, B.~D. Storey, M.~H. Oddy, C.-H. Chen, and J.~G. Santiago.
\newblock Instability of electrokinetic microchannel flows with conductivity
  gradients.
\newblock {\em Physics of Fluids}, 16(6):1922--1935, 2004.

\bibitem{Diffusion_InfiniteDilution}
X.~Liu, A.~Bardow, and T.~J.~H. Vlugt.
\newblock Multicomponent maxwell- stefan diffusivities at infinite dilution.
\newblock {\em Industrial \& Engineering Chemistry Research}, 50(8):4776--4782,
  2011.

\bibitem{MS_diffusion_ionic}
X.~Liu, T.~J.~H. Vlugt, and A.~Bardow.
\newblock {Maxwell--Stefan Diffusivities in Binary Mixtures of Ionic Liquids
  with Dimethyl Sulfoxide (DMSO) and H2O}.
\newblock {\em The Journal of Physical Chemistry B}, 115(26):8506--8517, 2011.

\bibitem{ZeroMachCombustion}
A.~Majda and J.~Sethian.
\newblock {The derivation and numerical solution of the equations for zero Mach
  number combustion}.
\newblock {\em Combustion science and technology}, 42(3):185--205, 1985.

\bibitem{newman2012electrochemical}
J.~Newman and K.~E. Thomas-Alyea.
\newblock {\em Electrochemical Systems}.
\newblock Wiley, 2012.

\bibitem{LowMachImplicit}
A.~J. Nonaka, Y.~Sun, J.~B. Bell, and A.~Donev.
\newblock {Low Mach Number Fluctuating Hydrodynamics of Binary Liquid
  Mixtures}.
\newblock {\em Communications in Applied Mathematics and Computational
  Science}, 10(2):163--204, 2015.

\bibitem{Posner2012}
J.~D. Posner, C.~L. Pérez, and J.~G. Santiago.
\newblock Electric fields yield chaos in microflows.
\newblock {\em Proceedings of the National Academy of Sciences},
  109(36):14353--14356, 2012.

\bibitem{robinson2012electrolyte}
R.~A. Robinson and R.~H. Stokes.
\newblock {\em Electrolyte Solutions: Second Revised Edition}.
\newblock Dover Books on Chemistry Series. Dover Publications, Incorporated,
  2012.

\bibitem{StagerredFluct_Inhomogeneous}
B.~Z. Shang, N.~K. Voulgarakis, and J.-W. Chu.
\newblock {Fluctuating hydrodynamics for multiscale simulation of inhomogeneous
  fluids: Mapping all-atom molecular dynamics to capillary waves}.
\newblock {\em J. Chem. Phys.}, 135:044111, 2011.

\bibitem{LLNS_Staggered}
F.~Balboa Usabiaga, J.~B. Bell, R.~Delgado-Buscalioni, A.~Donev, T.~G. Fai,
  B.~E. Griffith, and C.~S. Peskin.
\newblock {Staggered Schemes for Fluctuating Hydrodynamics}.
\newblock {\em SIAM J. Multiscale Modeling and Simulation}, 10(4):1369--1408,
  2012.

\bibitem{GiantFluctuations_Summary}
A.~Vailati, R.~Cerbino, S.~Mazzoni, M.~Giglio, C.~J. Takacs, and D.~S. Cannell.
\newblock {Gradient-driven fluctuations in microgravity.}
\newblock {\em Journal of physics. Condensed matter}, 24(28):284134, 2012.

\bibitem{GiantFluctuations_Nature}
A.~Vailati and M.~Giglio.
\newblock {Giant fluctuations in a free diffusion process}.
\newblock {\em Nature}, 390(6657):262--265, 1997.

\bibitem{NumericalElectrodynamicsBook}
U.~van Rienen.
\newblock {\em Numerical Methods in Computational Electrodynamics: Linear
  Systems in Practical Applications}.
\newblock Lecture Notes in Computational Science and Engineering. Springer
  Berlin Heidelberg, 2012.

\bibitem{Electrolytes_DH_review}
Luis~M Varela, Manuel Garcia, and Victor Mosquera.
\newblock Exact mean-field theory of ionic solutions: non-debye screening.
\newblock {\em Physics reports}, 382(1):1--111, 2003.

\bibitem{FluctHydroNonEq_Book}
J.~M.~O.~De Zarate and J.~V. Sengers.
\newblock {\em {Hydrodynamic fluctuations in fluids and fluid mixtures}}.
\newblock Elsevier Science Ltd, 2007.

\bibitem{LatticeBoltzElectrolytes}
J.~Zudrop, S.~Roller, and P.~Asinari.
\newblock {Lattice Boltzmann scheme for electrolytes by an extended
  Maxwell-Stefan approach}.
\newblock {\em Phys. Rev. E}, 89:053310, May 2014.

\end{thebibliography}

\end{document}